\documentclass{IEEEtran}
\usepackage{cite}
\usepackage{amsmath,amssymb,amsfonts}
\usepackage{graphicx,color}
\usepackage{textcomp}
\usepackage{xcolor}
\usepackage{hyperref}
\hypersetup{hidelinks=true}
\usepackage{subcaption}
\def\BibTeX{{\rm B\kern-.05em{\sc i\kern-.025em b}\kern-.08em
    T\kern-.1667em\lower.7ex\hbox{E}\kern-.125emX}}
\AtBeginDocument{\definecolor{tmlcncolor}{cmyk}{0.93,0.59,0.15,0.02}\definecolor{NavyBlue}{RGB}{0,86,125}}
\newcommand{\cdfaxiswidth}{\dimexpr\columnwidth-1.2cm\relax}

\usepackage{multirow}
\usepackage{algpseudocode}
\usepackage[nolist,nohyperlinks]{acronym}
\usepackage{bm}
\usepackage{microtype}
\usepackage{booktabs}
\usepackage{siunitx}

\usepackage{tikz}
\usetikzlibrary{%
    angles,%
    arrows,%
    arrows.meta,%
    babel,%
    backgrounds,%
    calc,%
    fit,%
    matrix,%
    mindmap,%
    patterns,%
    positioning,%
    quotes,%
    shapes,%
    shadows,%
    tikzmark,%
    trees,%
}
\usepackage{pgfplots}
\usepgfplotslibrary{groupplots}
\usepgfplotslibrary{statistics}
\pgfplotsset{width=\textwidth*0.4,compat=1.9}

\usepackage{xcolor}

\acrodef{DT}[DT]{digital twin}
\acrodef{RAN}[RAN]{radio access network}
\acrodef{GAN}[GAN]{generative adversarial network}
\acrodef{GMM}[GMM]{Gaussian mixture model}
\acrodef{HMM}[HMM]{hidden Markov model}
\acrodef{IAT}[IAT]{inter arrival time}
\acrodef{MDN}[MDN]{mixture density network}
\acrodef{EM}[EM]{expectation-maximization}
\acrodef{HTTP}[HTTP]{hypertext transfer protocol}
\acrodef{UDP}[UDP]{datagram protocol}
\acrodef{MTU}[MTU]{Ethernet maximum transmission unit}
\acrodef{NLL}[NLL]{negative log likelihood}
\acrodef{GRU}[GRU]{gated recurrent unit}
\acrodef{CDF}[CDF]{cumulative distribution function}
\acrodef{KL}[KL]{Kullback-Leibler}
\acrodef{KS}[KS]{Kolmogorov-Smirnov}
\acrodef{PSD}[PSD]{power spectral density}
\acrodef{ACF}[ACF]{autocorrelation function}
\acrodef{AC}[AC]{autocorrelation}
\acrodef{WD}[WD]{Wasserstein distance}
\acrodef{UE}[UE]{user equipment}
\acrodef{CPU}[CPU]{central processing unit}
\acrodef{PMF}[PMF]{probability mass functions}
\acrodef{MAC}[MAC]{medium access control}
\acrodef{FTPS}[FTPS]{file transfer protocol secure}

\renewcommand{\textcolor}[2]{#2}  

\begin{document}

\definecolor{kit-green}{RGB}{0, 150, 130}
\colorlet{KITgreen}{kit-green}
\definecolor{kit-green100}{RGB}{0, 150, 130}
\definecolor{kit-green90}{rgb}{0.1, 0.6294, 0.5588}
\definecolor{kit-green80}{rgb}{0.2, 0.6706, 0.6078}
\definecolor{kit-green75}{rgb}{0.25, 0.6912, 0.6324}
\definecolor{kit-green70}{rgb}{0.3, 0.7118, 0.6569}
\definecolor{kit-green60}{rgb}{0.4, 0.7529, 0.7059}
\definecolor{kit-green50}{rgb}{0.5, 0.7941, 0.7549}
\definecolor{kit-green40}{rgb}{0.6, 0.8353, 0.8039}
\definecolor{kit-green30}{rgb}{0.7, 0.8765, 0.8529}
\definecolor{kit-green25}{rgb}{0.75, 0.8971, 0.8775}
\definecolor{kit-green20}{rgb}{0.8, 0.9176, 0.902}
\definecolor{kit-green15}{rgb}{0.85, 0.9382, 0.9265}
\definecolor{kit-green10}{rgb}{0.9, 0.9588, 0.951}
\definecolor{kit-green5}{rgb}{0.95, 0.9794, 0.9755}

\definecolor{kit-blue}{RGB}{70, 100, 170}
\colorlet{KITblue}{kit-blue}
\definecolor{kit-blue100}{RGB}{70, 100, 170}
\definecolor{kit-blue90}{rgb}{0.3471, 0.4529, 0.7}
\definecolor{kit-blue80}{rgb}{0.4196, 0.5137, 0.7333}
\definecolor{kit-blue75}{rgb}{0.4559, 0.5441, 0.75}
\definecolor{kit-blue70}{rgb}{0.4922, 0.5745, 0.7667}
\definecolor{kit-blue60}{rgb}{0.5647, 0.6353, 0.8}
\definecolor{kit-blue50}{rgb}{0.6373, 0.6961, 0.8333}
\definecolor{kit-blue40}{rgb}{0.7098, 0.7569, 0.8667}
\definecolor{kit-blue30}{rgb}{0.7824, 0.8176, 0.9}
\definecolor{kit-blue25}{rgb}{0.8186, 0.848, 0.9167}
\definecolor{kit-blue20}{rgb}{0.8549, 0.8784, 0.9333}
\definecolor{kit-blue15}{rgb}{0.8912, 0.9088, 0.95}
\definecolor{kit-blue10}{rgb}{0.9275, 0.9392, 0.9667}
\definecolor{kit-blue5}{rgb}{0.9637, 0.9696, 0.9833}

\definecolor{kit-red}{RGB}{162, 34, 35}
\colorlet{KITred}{kit-red}
\definecolor{kit-red100}{RGB}{162, 34, 35}
\definecolor{kit-red90}{rgb}{0.6718, 0.22, 0.2235}
\definecolor{kit-red80}{rgb}{0.7082, 0.3067, 0.3098}
\definecolor{kit-red75}{rgb}{0.7265, 0.35, 0.3529}
\definecolor{kit-red70}{rgb}{0.7447, 0.3933, 0.3961}
\definecolor{kit-red60}{rgb}{0.7812, 0.48, 0.4824}
\definecolor{kit-red50}{rgb}{0.8176, 0.5667, 0.5686}
\definecolor{kit-red40}{rgb}{0.8541, 0.6533, 0.6549}
\definecolor{kit-red30}{rgb}{0.8906, 0.74, 0.7412}
\definecolor{kit-red25}{rgb}{0.9088, 0.7833, 0.7843}
\definecolor{kit-red20}{rgb}{0.9271, 0.8267, 0.8275}
\definecolor{kit-red15}{rgb}{0.9453, 0.87, 0.8706}
\definecolor{kit-red10}{rgb}{0.9635, 0.9133, 0.9137}
\definecolor{kit-red5}{rgb}{0.9818, 0.9567, 0.9569}

\definecolor{kit-yellow}{RGB}{252, 229, 0}
\colorlet{KITyellow}{kit-yellow}
\definecolor{kit-yellow100}{RGB}{252, 229, 0}
\definecolor{kit-yellow90}{rgb}{0.9894, 0.9082, 0.1}
\definecolor{kit-yellow80}{rgb}{0.9906, 0.9184, 0.2}
\definecolor{kit-yellow75}{rgb}{0.9912, 0.9235, 0.25}
\definecolor{kit-yellow70}{rgb}{0.9918, 0.9286, 0.3}
\definecolor{kit-yellow60}{rgb}{0.9929, 0.9388, 0.4}
\definecolor{kit-yellow50}{rgb}{0.9941, 0.949, 0.5}
\definecolor{kit-yellow40}{rgb}{0.9953, 0.9592, 0.6}
\definecolor{kit-yellow30}{rgb}{0.9965, 0.9694, 0.7}
\definecolor{kit-yellow25}{rgb}{0.9971, 0.9745, 0.75}
\definecolor{kit-yellow20}{rgb}{0.9976, 0.9796, 0.8}
\definecolor{kit-yellow15}{rgb}{0.9982, 0.9847, 0.85}
\definecolor{kit-yellow10}{rgb}{0.9988, 0.9898, 0.9}
\definecolor{kit-yellow5}{rgb}{0.9994, 0.9949, 0.95}

\definecolor{kit-orange}{RGB}{223, 155, 27}
\colorlet{KITorange}{kit-orange}
\definecolor{kit-orange100}{RGB}{223, 155, 27}
\definecolor{kit-orange90}{rgb}{0.8871, 0.6471, 0.1953}
\definecolor{kit-orange80}{rgb}{0.8996, 0.6863, 0.2847}
\definecolor{kit-orange75}{rgb}{0.9059, 0.7059, 0.3294}
\definecolor{kit-orange70}{rgb}{0.9122, 0.7255, 0.3741}
\definecolor{kit-orange60}{rgb}{0.9247, 0.7647, 0.4635}
\definecolor{kit-orange50}{rgb}{0.9373, 0.8039, 0.5529}
\definecolor{kit-orange40}{rgb}{0.9498, 0.8431, 0.6424}
\definecolor{kit-orange30}{rgb}{0.9624, 0.8824, 0.7318}
\definecolor{kit-orange25}{rgb}{0.9686, 0.902, 0.7765}
\definecolor{kit-orange20}{rgb}{0.9749, 0.9216, 0.8212}
\definecolor{kit-orange15}{rgb}{0.9812, 0.9412, 0.8659}
\definecolor{kit-orange10}{rgb}{0.9875, 0.9608, 0.9106}
\definecolor{kit-orange5}{rgb}{0.9937, 0.9804, 0.9553}

\definecolor{kit-lightgreen}{RGB}{140, 182, 60}
\colorlet{KITlightgreen}{kit-lightgreen}
\definecolor{kit-lightgreen100}{RGB}{140, 182, 60}
\definecolor{kit-lightgreen90}{rgb}{0.5941, 0.7424, 0.3118}
\definecolor{kit-lightgreen80}{rgb}{0.6392, 0.771, 0.3882}
\definecolor{kit-lightgreen75}{rgb}{0.6618, 0.7853, 0.4265}
\definecolor{kit-lightgreen70}{rgb}{0.6843, 0.7996, 0.4647}
\definecolor{kit-lightgreen60}{rgb}{0.7294, 0.8282, 0.5412}
\definecolor{kit-lightgreen50}{rgb}{0.7745, 0.8569, 0.6176}
\definecolor{kit-lightgreen40}{rgb}{0.8196, 0.8855, 0.6941}
\definecolor{kit-lightgreen30}{rgb}{0.8647, 0.9141, 0.7706}
\definecolor{kit-lightgreen25}{rgb}{0.8873, 0.9284, 0.8088}
\definecolor{kit-lightgreen20}{rgb}{0.9098, 0.9427, 0.8471}
\definecolor{kit-lightgreen15}{rgb}{0.9324, 0.9571, 0.8853}
\definecolor{kit-lightgreen10}{rgb}{0.9549, 0.9714, 0.9235}
\definecolor{kit-lightgreen5}{rgb}{0.9775, 0.9857, 0.9618}

\definecolor{kit-purple}{RGB}{163, 16, 124}
\colorlet{KITpurple}{kit-purple}
\definecolor{kit-purple100}{RGB}{163, 16, 124}
\definecolor{kit-purple90}{rgb}{0.6753, 0.1565, 0.5376}
\definecolor{kit-purple80}{rgb}{0.7114, 0.2502, 0.589}
\definecolor{kit-purple75}{rgb}{0.7294, 0.2971, 0.6147}
\definecolor{kit-purple70}{rgb}{0.7475, 0.3439, 0.6404}
\definecolor{kit-purple60}{rgb}{0.7835, 0.4376, 0.6918}
\definecolor{kit-purple50}{rgb}{0.8196, 0.5314, 0.7431}
\definecolor{kit-purple40}{rgb}{0.8557, 0.6251, 0.7945}
\definecolor{kit-purple30}{rgb}{0.8918, 0.7188, 0.8459}
\definecolor{kit-purple25}{rgb}{0.9098, 0.7657, 0.8716}
\definecolor{kit-purple20}{rgb}{0.9278, 0.8125, 0.8973}
\definecolor{kit-purple15}{rgb}{0.9459, 0.8594, 0.9229}
\definecolor{kit-purple10}{rgb}{0.9639, 0.9063, 0.9486}
\definecolor{kit-purple5}{rgb}{0.982, 0.9531, 0.9743}

\definecolor{kit-brown}{RGB}{167, 130, 46}
\colorlet{KITbrown}{kit-brown}
\definecolor{kit-brown100}{RGB}{167, 130, 46}
\definecolor{kit-brown90}{rgb}{0.6894, 0.5588, 0.2624}
\definecolor{kit-brown80}{rgb}{0.7239, 0.6078, 0.3443}
\definecolor{kit-brown75}{rgb}{0.7412, 0.6324, 0.3853}
\definecolor{kit-brown70}{rgb}{0.7584, 0.6569, 0.4263}
\definecolor{kit-brown60}{rgb}{0.7929, 0.7059, 0.5082}
\definecolor{kit-brown50}{rgb}{0.8275, 0.7549, 0.5902}
\definecolor{kit-brown40}{rgb}{0.862, 0.8039, 0.6722}
\definecolor{kit-brown30}{rgb}{0.8965, 0.8529, 0.7541}
\definecolor{kit-brown25}{rgb}{0.9137, 0.8775, 0.7951}
\definecolor{kit-brown20}{rgb}{0.931, 0.902, 0.8361}
\definecolor{kit-brown15}{rgb}{0.9482, 0.9265, 0.8771}
\definecolor{kit-brown10}{rgb}{0.9655, 0.951, 0.918}
\definecolor{kit-brown5}{rgb}{0.9827, 0.9755, 0.959}

\definecolor{kit-cyan}{RGB}{35, 161, 224}
\colorlet{KITcyan}{kit-cyan}
\colorlet{KITcyanblue}{kit-cyan}
\definecolor{kit-cyan100}{RGB}{35, 161, 224}
\definecolor{kit-cyan90}{rgb}{0.2235, 0.6682, 0.8906}
\definecolor{kit-cyan80}{rgb}{0.3098, 0.7051, 0.9027}
\definecolor{kit-cyan75}{rgb}{0.3529, 0.7235, 0.9088}
\definecolor{kit-cyan70}{rgb}{0.3961, 0.742, 0.9149}
\definecolor{kit-cyan60}{rgb}{0.4824, 0.7788, 0.9271}
\definecolor{kit-cyan50}{rgb}{0.5686, 0.8157, 0.9392}
\definecolor{kit-cyan40}{rgb}{0.6549, 0.8525, 0.9514}
\definecolor{kit-cyan30}{rgb}{0.7412, 0.8894, 0.9635}
\definecolor{kit-cyan25}{rgb}{0.7843, 0.9078, 0.9696}
\definecolor{kit-cyan20}{rgb}{0.8275, 0.9263, 0.9757}
\definecolor{kit-cyan15}{rgb}{0.8706, 0.9447, 0.9818}
\definecolor{kit-cyan10}{rgb}{0.9137, 0.9631, 0.9878}
\definecolor{kit-cyan5}{rgb}{0.9569, 0.9816, 0.9939}

\definecolor{kit-gray}{RGB}{0, 0, 0}
\colorlet{KITgray}{kit-gray}
\definecolor{kit-gray100}{RGB}{0, 0, 0}
\definecolor{kit-gray90}{rgb}{0.1, 0.1, 0.1}
\definecolor{kit-gray80}{rgb}{0.2, 0.2, 0.2}
\definecolor{kit-gray75}{rgb}{0.25, 0.25, 0.25}
\definecolor{kit-gray70}{rgb}{0.3, 0.3, 0.3}
\definecolor{kit-gray60}{rgb}{0.4, 0.4, 0.4}
\definecolor{kit-gray50}{rgb}{0.5, 0.5, 0.5}
\definecolor{kit-gray40}{rgb}{0.6, 0.6, 0.6}
\definecolor{kit-gray30}{rgb}{0.7, 0.7, 0.7}
\definecolor{kit-gray25}{rgb}{0.75, 0.75, 0.75}
\definecolor{kit-gray20}{rgb}{0.8, 0.8, 0.8}
\definecolor{kit-gray15}{rgb}{0.85, 0.85, 0.85}
\definecolor{kit-gray10}{rgb}{0.9, 0.9, 0.9}
\definecolor{kit-gray5}{rgb}{0.95, 0.95, 0.95}

\definecolor{KITpalegreen}{RGB}{130,190,60}
\colorlet{kit-maigreen100}{KITpalegreen}
\colorlet{kit-maigreen70}{KITpalegreen!70}
\colorlet{kit-maigreen50}{KITpalegreen!50}
\colorlet{kit-maigreen30}{KITpalegreen!30}
\colorlet{kit-maigreen15}{KITpalegreen!15}


\title{Packet-Level Traffic Modeling with Heavy-Tailed Payload and Inter-Arrival Distributions for Digital Twins}

\author{Enes Koktas and Peter Rost 
		\thanks{This research was supported by the German Federal Ministry of Research, Technology and Space (BMFTR) under grant number 16KIS2259 (SUSTAINET-inNOvAte).}
        }

\maketitle

\begin{abstract}
Digital twins of radio access networks require packet-level traffic generators that reproduce the size and timing of packets while remaining compact and easy to \textcolor{red}{configure} as traffic changes. We address this need with a hybrid generator that combines a small hidden Markov model, which captures buffering, streaming, and idle states, with a mixture density network that models the joint distribution of payload length and \ac{IAT} in each state using Student-t mixtures. The state space and emission family are designed to handle heavy-tailed \acp{IAT} by anchoring an explicit idle state in the tail and allowing each component to adapt its tail thickness. We evaluate the model on public traces of smart home and encrypted media traffic and compare it with recent neural network and transformer based generators as well as hidden Markov baselines. Across four datasets and metrics, including average per-flow cumulative distribution functions, autocorrelation based measures of temporal structure, and Wasserstein distances, the proposed generator matches the real traffic most closely in the majority of cases while using orders of magnitude fewer parameters. \textcolor{red}{The full model occupies around 0.2 MB and achieves a generation throughput of over 21,000 packets per second in our experiments, making it suitable for deployment inside digital twins where minimal memory footprint and low-latency execution are critical.}
\end{abstract}

\begin{IEEEkeywords}
Heavy tail, Hidden Markov model, Mixture density network, Mobile communication, Network digital twin, Synthetic traffic generation
\end{IEEEkeywords}

\maketitle

\section{INTRODUCTION}\label{sec:Introduction}
\IEEEPARstart{I}{n} recent years, \acp{DT} have received more attention because they enable optimization, autonomy, predictive maintenance, real-time monitoring, root cause analysis, problem detection, and risk-free what-if scenario analysis for present and future communication protocols and operations. A \ac{DT} framework consists of three main components: the real space, the virtual space, and a link that allows communication between the two \cite{9899718}.  The virtual space replicates the physical counterpart in real-time. Existing network operations are often already automated but need feedback from an operating network, in which \acp{DT} may play a major part. Various use cases of \acp{DT} at different stages of a communication system require different resolution of available statistics. Exemplary use cases include handover optimization, resource allocation, network planning, and traffic offloading. A network \ac{DT} can combine data collection and processing, and AI modeling to examine the network's current condition, forecast future patterns, and facilitate predictive maintenance \cite{irtf-nmrg}.

\begin{table*}[!ht]
\centering
\caption{\textcolor{red}{Comparison of State-of-the-Art Traffic Generation Approaches for Digital Twins.}}
\label{tab:sota_comparison}
    \begin{tabular}{lll}
        \hline
        \textbf{Approach} & \textbf{Advantages} & \textbf{Disadvantages} \\ \hline
        \begin{tabular}[c]{@{}l@{}}Analytical Models \\ (e.g., Poisson, Markov)\end{tabular} &
          \begin{tabular}[c]{@{}l@{}}Computationally trivial, real-time execution, zero\\ memory footprint\end{tabular} &
          \begin{tabular}[c]{@{}l@{}}Fails to capture modern heavy-tailed distributions\\ and bursty traffic behaviors\end{tabular} \\ \hline
        \begin{tabular}[c]{@{}l@{}}Trace Replay \\ (e.g., TRex, Harpoon)\end{tabular} &
          \begin{tabular}[c]{@{}l@{}}Perfect realism for the specific captured network\\ scenario\end{tabular} &
          \begin{tabular}[c]{@{}l@{}}Massive storage overhead, low adaptability to\\ changing network states\end{tabular} \\ \hline
        \begin{tabular}[c]{@{}l@{}}GANs \& VAEs \\ (e.g., TimeGAN)\end{tabular} &
          \begin{tabular}[c]{@{}l@{}}High distributional fidelity, learns complex temporal\\ dependencies directly from data\end{tabular} &
          \begin{tabular}[c]{@{}l@{}}Unstable adversarial training (mode collapse), \\ typically requires large parameter counts\end{tabular} \\ \hline
        \begin{tabular}[c]{@{}l@{}}Diffusion Models \\ (e.g., NetDiffusion)\end{tabular} &
          \begin{tabular}[c]{@{}l@{}}Exceptionally high fidelity, stable training,\\ protocol-aware synthesis\end{tabular} &
          \begin{tabular}[c]{@{}l@{}}High inference latency due to iterative denoising,\\ prohibitive for live, edge-native \acp{DT}\end{tabular} \\ \hline
        \begin{tabular}[c]{@{}l@{}}Proposed Model \\ (HMM + MDN)\end{tabular} &
          \begin{tabular}[c]{@{}l@{}}Real-time inference, compact footprint ($\sim$0.2 MB),\\ captures heavy-tails\end{tabular} &
          \begin{tabular}[c]{@{}l@{}}Explicit state modeling may require structural\\ retuning if fundamental application behaviors\\ drastically change\end{tabular} \\ \hline
    \end{tabular}%
\end{table*}

\subsection{Problem definition}
A network \ac{DT} that represents a \ac{RAN} should reproduce how traffic is offered to the network at packet level, not only at a coarse rate or volume level, because schedulers, link adaptation, buffer management, and congestion control all depend on the timing and size of packets. Real packet traces are often heavy tailed, meaning they are characterized by rare but large \ac{IAT}, where the measurements on wired and wireless links show heavy-tailed waiting times and session durations \cite{650143}. In our datasets, a small fraction of packets are separated by gaps that are orders of magnitude longer than the median, which influences whether buffers drain, inactivity timers expire, and flows return from idle. If a generator underestimates the frequency or length of such gaps, the \ac{DT} may underestimate delay, queue occupancy, and timeout probability, and thus provide overly optimistic performance predictions. At the same time, streaming full packet traces from the physical network to the virtual space and retraining a large model whenever conditions change is not practical due to data security risks, limited backbone capacity, and additional overhead. The problem we address in this work is therefore to design a compact and interpretable packet generator that can be trained on an initial dataset that captures the heavy-tailed structure of payload size and \acp{IAT}
so that the \ac{DT} can follow traffic changes without requiring continuous transfer of raw traces.

\subsection{Related Work}\label{subsec:related_work}
Early synthetic traffic generators such as TRex, NS-3, and Harpoon combine packet or flow replay with simulation models that are configured by hand \cite{3673792}. They are practical for reproducing known scenarios but rely on user defined templates that must be tuned by experts for each test case. As a result they struggle to match the detailed behavior of packet sizes and \acp{IAT} observed in modern traces. 
Recent work has therefore turned to data driven generative models that learn directly from packet traces. Several articles apply \acp{GAN} to synthesize network flows, with recurrent layers in the generator and discriminator to better capture temporal patterns \cite{3423643,3664655}. \textcolor{red}{To specifically address the temporal dynamics in sequential data, TimeGAN \cite{NEURIPS2019_c9efe5f2} extends the adversarial approach by incorporating a supervised learning objective. This explicitly guides the network to accurately reproduce the step-by-step transition patterns of the original traces.} Hybrid schemes combine autoencoders with Wasserstein GANs to operate in a latent space, for example by encoding Internet of Things packet size sequences, sampling new latent codes with a \ac{GAN}, and decoding them into synthetic flows \cite{9320384}. NeCSTGen takes a similar latent approach: a variational autoencoder compresses packets, a Gaussian mixture and recurrent neural network model the cluster transitions, and the resulting chain is used to produce packet, flow, and aggregate level traces \cite{10000731}. These neural generators can reproduce a broad class of marginal statistics, but adversarial training is sensitive to initialization and loss design.

\textcolor{red}{To overcome these training instabilities, recent works have explored diffusion models for network traffic synthesis. For instance, NetDiffusion \cite{netdiffusion} leverages protocol-aware diffusion processes to generate high-fidelity packet captures, while STOUTER \cite{stouter} employs spatio-temporal diffusion for cellular traffic demand modeling. While these models achieve exceptional distributional realism, their iterative denoising algorithms incur substantial computational overhead and inference latency. This computational burden makes them difficult to deploy as live, continuous traffic nodes at the network edge within a real-time \ac{DT}.} 
\textcolor{red}{To clearly position our work within this landscape, we have summarized the advantages and disadvantages of state-of-the-art approaches in Table \ref{tab:sota_comparison}.}

\textcolor{red}{More recently, transformer-based sequence models have been proposed for network traffic analysis and generation. For instance, the pre-trained transformer in \cite{qu2024trafficgpt} utilizes a linear attention mechanism to process raw byte-level packets over extended contexts. While such architectures are highly effective when massive centralized datasets and vast compute resources are available, the computational cost of byte-level autoregression and their large memory footprint make them difficult to deploy and fine-tune on edge hardware for dynamic, local \ac{RAN} \acp{DT}.}
In \cite{9145646}, the authors fit two independent \acp{HMM}, one for payload length and one for \ac{IAT}, to video streaming and gaming traces using the Baum-Welch algorithm. This captures the one dimensional marginals of size and timing but might miss their joint dependence, which is important for buffer dynamics and delay.

\subsection{Contribution}
In this article we propose a compact packet level traffic generator for network \acp{DT} that factorizes each flow into a small hidden state process and a \ac{MDN}. The \ac{HMM} captures coarse states such as buffering, steady streaming, and idle, while the \ac{MDN} learns the joint distribution of payload length and \ac{IAT} within each state. The design is explicitly tailored to heavy tailed \acp{IAT}. We augment the state space with an idle state that is anchored in the tail of the \ac{IAT} distribution and replace Gaussian components by Student-t kernels so that rare long gaps and bursts receive appropriate probability mass. \textcolor{red}{Unlike state-of-the-art neural and transformer models with large memory footprints, our hybrid generator provides a lightweight operational cornerstone for network \acp{DT}. By utilizing a structured probabilistic architecture, it achieves competitive distributional fidelity, flow diversity, and realistic heavy tails while keeping the memory footprint relatively low. This ensures practical, low-overhead deployment on resource-constrained edge hardware.}

\section{Dataset and Notation}\label{sec:dataset_notation}
We evaluate the proposed generator on four packet level traces that cover file transfers, smart home traffic, and encrypted media streaming. The first trace contains \ac{UDP} traffic \textcolor{red}{recorded from a Google Home device} \cite{10000731}. The other three traces contain Facebook audio, Facebook video, and \ac{FTPS} file transfer sessions from the virtual private network traffic dataset described in \cite{draper2016}. In all four datasets we keep the original flow identifiers and retain only the fields needed by our model, namely payload length and \ac{IAT} for each packet. Table~\ref{tab:dataset_summary} summarizes the main characteristics of these traces in terms of number of flows, total packets, packet count per flow, volume, and average flow duration. The UDP traces are small but contain many short flows, while the Facebook audio, Facebook video, and \ac{FTPS} traces have fewer flows with longer sessions and heavier tails.

\begin{table}[]
\centering
\caption{Summary statistics of the datasets.}
\label{tab:dataset_summary}
\resizebox{\columnwidth}{!}{%
\begin{tabular}{lllll}
\hline
Feature Name & \begin{tabular}[c]{@{}l@{}}Google Home\\ (UDP)\end{tabular} & \begin{tabular}[c]{@{}l@{}}Facebook\\ Audio\end{tabular} & \begin{tabular}[c]{@{}l@{}}Facebook\\ Video\end{tabular} & \ac{FTPS} \\ \hline
File Size (MB) & $0.30$ & $2.4$ & $4.4$ & $180.03$ \\
Total Flows & $588$ & $250$ & $173$ & $585$ \\
Total Packets & $10{,}000$ & $91{,}737$ & $156{,}982$ & $5{,}743{,}492$ \\
Average Pkts./Flow & $17.0$ & $366.9$ & $907.4$ & $9{,}817.9$ \\
Total Volume (\SI{}{MB}) & $3.91$ & $10.57$ & $162.67$ & $5{,}291.1$ \\
\begin{tabular}[c]{@{}l@{}}Average Flow\\ Duration (\SI{}{\second})\end{tabular} & $8.87$ & $97.36$ & $55.78$ & $91.17$ \\ \hline
\end{tabular}%
}
\end{table}
We work at packet level and represent each captured flow as an ordered sequence of packet tuples. The raw dataset consists of one row per packet in the form \textit{flow index, payload length, time difference}, where flow index is an identifier that groups packets into flows, payload length is the application payload in bytes, and time difference is the \ac{IAT} between consecutive packets in seconds.

Let $\mathbb{R}_{\ge 0} = \{x \in \mathbb{R} \mid x \ge 0\}$ denote the set of non-negative real numbers. For a given flow index $i$ with length $L_i \in \mathbb{N}$, the $t$-th packet \textcolor{red}{in the sequence (where $t$ serves as a discrete sequence index} is denoted by the vector $\bm{x}_{i,t} \in \mathbb{R}_{\ge 0}^2$ with
\begin{equation} \label{eq:packet_def}
\bm{x}_{i,t} = [p_{i,t}, \delta_{i,t}]^\top,
\end{equation}
where $p_{i,t}$ is the payload length in bytes and $\delta_{i,t}$ is the \ac{IAT} in seconds, defined as the time interval between the arrival of the current packet $t$ and the previous packet $t-1$, \textcolor{red}{where $\delta_{i,1} = 0$ for the first packet of a flow}.

To make the statistics well-conditioned while preserving the heavy right tail of \acp{IAT} that motivated our model design, we cap only unrealistically large gaps during preprocessing used to fit scalers (e.g., one hour for \acp{IAT}), and bound payloads to the valid \ac{MTU} range seen in the traces.

\textcolor{red}{
For learning, we apply a monotone normalizing transform componentwise, where $\ln(\cdot)$ denotes the natural logarithm, and then normalize with moments estimated on training flows only. The normalized packet is
\begin{equation}\label{eq:normalized_packet}
\bm{z}_{i,t} = \left( \frac{\ln(1+p_{i,t}) - m_p}{s_p}, \frac{\ln(1+\delta_{i,t}) - m_\delta}{s_\delta} \right)^\top,
\end{equation}
where $m_p, m_\delta$ and $s_p, s_\delta$ are the mean and standard deviation computed over all packets in the training flows.
}
Let $\ell_i = \ln(L_i)$ be the log transformed flow length, and let $m_\ell$ and $s_\ell$ be its training mean and standard deviation; we define the scalar
\begin{equation}\label{eq:norm_flow_length}
\xi_i = \frac{\ell_i - m_\ell}{s_\ell},
\end{equation}
for each flow $i$. 
We split the dataset into training and testing sets by flow to preserve the flow structure. 
During synthesis, each test flow keeps its original identifier and length, which enables paired, flow-wise evaluations between real and synthetic sequences.

The dataset exhibits heavy-tailed \ac{IAT}. We quantify this tail by computing a tail threshold $\tau_\delta \in \mathbb{R}_{+}$ defined as the empirical $99.8$th percentile of the $\ln(\delta_{i,t}+1)$ over all packets in the training set. This scalar is used to define an “idle” state in the state space when required. We normalize $\tau_\delta$ through
\begin{equation}\label{eq:normalized_threshold}
\upsilon_{\delta}=\frac{\tau_{\delta}-m_\delta}{s_\delta}.
\end{equation}

For state-related notation, we use $K \in \mathbb{N}$ for the number of latent states and 
$k \in \{1,\ldots,K\}$ for the index of a latent state. The latent state is
$q_{i,t} \in \{1,\ldots,K\}$, and
\begin{equation}\label{eq:one_hot_state}
\bm{e}^{(k)} \in \{0,1\}^K, \quad \text{with } e_j^{(k)} = \begin{cases} 1 & \text{if } j = k, \\ 0 & \text{otherwise}, \end{cases}
\end{equation}
denotes the one-hot vector of state $k$.
We follow this section’s notations in the remainder of the manuscript when specifying the \ac{HMM}, idle state augmentation, and the \ac{MDN}.

\textcolor{red}{
\section{Problem statement}\label{subsec:formal_problem}
Let the training set be
\begin{equation}
\mathcal{D}_{\mathrm{tr}}
=
\left\{
\bm{X}^{(i)}
\right\}_{i=1}^{N_{\mathrm{tr}}},
\end{equation}
where $N_{\mathrm{tr}} \in \mathbb{N}$ is the number of training flows and
\begin{equation}
\bm{X}^{(i)}
=
\left(
\bm{x}_{i,1},
\bm{x}_{i,2},
\ldots,
\bm{x}_{i,L_i}
\right)
\end{equation}
is the $i$-th flow in the training data. }

\textcolor{red}{The objective is to learn a compact generator with parameters $\bm{\Omega} = \{\bm{\Theta},\bm{\varphi}\}$, where $\bm{\Theta}$ and $\bm{\varphi}$ denote the parameters of the \ac{HMM} and \ac{MDN}, respectively. While the generator inherently operates in the normalized continuous space to produce synthetic latent sequences $\widehat{\bm{Z}}^{(i)} \sim p_{\bm{\Omega}}(\cdot \mid \xi_i)$, these are deterministically inverse-transformed to yield the generated flow $\widehat{\bm{X}}^{(i)}$. The overarching goal is to optimize $\bm{\Omega}$ such that $\widehat{\bm{X}}^{(i)}$ is statistically similar to the corresponding real flow $\bm{X}^{(i)}$ and, at the dataset level, the synthetic set reproduces the same heavy-tailed traffic structure as the real set.}

\textcolor{red}{In this work, similarity is measured through three complementary properties. First, the generator should reproduce the packet-level marginal fidelity of payload length and \ac{IAT}. We evaluate this fidelity by calculating the sum of absolute differences between \acp{PMF} discretized into $B=1000$ logarithmically spaced bins. Let
\begin{equation}
\overline{P}_{p}^{\mathrm{r}}(b)
=
\frac{1}{N_{\mathrm{te}}}
\sum_{i=1}^{N_{\mathrm{te}}}
P_{p,i}^{\mathrm{r}}(b)
\end{equation}
and
\begin{equation}
\overline{P}_{p}^{\mathrm{g}}(b)
=
\frac{1}{N_{\mathrm{te}}}
\sum_{i=1}^{N_{\mathrm{te}}}
P_{p,i}^{\mathrm{g}}(b),
\end{equation}
where $P_{p,i}^{\mathrm{r}}(b)$ and $P_{p,i}^{\mathrm{g}}(b)$ denote the fraction of real and generated payload lengths of flow $i$ that fall into bin $b \in \{1, \dots, B\}$, representing the empirical per-flow PMFs and $N_{\mathrm{te}} \in \mathbb{N}$ is the number of test flows. We define $\overline{P}_{\delta}^{\mathrm{r}}(b)$ and $\overline{P}_{\delta}^{\mathrm{g}}(b)$ analogously for \acp{IAT}. The marginal fidelity discrepancy is then defined as a vector containing the total absolute difference for each feature independently:
\begin{equation}
\bm{\mathcal{D}}_{\mathrm{marg}}
=
\left[
\sum_{b=1}^{B} \left| \overline{P}_{p}^{\mathrm{r}}(b) - \overline{P}_{p}^{\mathrm{g}}(b) \right|, \quad
\sum_{b=1}^{B} \left| \overline{P}_{\delta}^{\mathrm{r}}(b) - \overline{P}_{\delta}^{\mathrm{g}}(b) \right|
\right]^{\top}\!,
\end{equation}
where the maximum possible discrepancy for each element is bounded by $2.0$.}

\textcolor{red}{Second, the generator should preserve short-range temporal dependence within each flow, as \ac{AC} has been used in traffic measurements to characterize dependence and burstiness over multiple time scales \cite{paxson1995poisson}. We quantify this by comparing the \ac{AC} at various lags, where a lag $w$ represents the distance in the sequence between two packets. Let $\overline{\rho}_{p}^{\mathrm{r}}(w)$ and $\overline{\rho}_{p}^{\mathrm{g}}(w)$ denote the empirical \ac{AC} of payload length at lag $w$, averaged over all real and generated flows, respectively. Computing this per-flow avoids artificially mixing traffic from distinct connections. Let $\overline{\rho}_{\delta}^{\mathrm{r}}(w)$ and $\overline{\rho}_{\delta}^{\mathrm{g}}(w)$ denote the corresponding average \ac{AC} values for \ac{IAT}. For a maximum lag $W \in \mathbb{N}$, the temporal discrepancy is defined as a vector containing the root-mean-square errors of the autocorrelations:
\begin{equation}
    \begin{aligned}
        \bm{\mathcal{D}}_{\mathrm{temp}} &= 
        \Bigg[
        \sqrt{ \frac{1}{W} \sum_{w}^{} \left( \overline{\rho}_{p}^{\mathrm{r}}(w) - \overline{\rho}_{p}^{\mathrm{g}}(w) \right)^2 }, \\
        &\qquad \sqrt{ \frac{1}{W} \sum_{w}^{} \left( \overline{\rho}_{\delta}^{\mathrm{r}}(w) - \overline{\rho}_{\delta}^{\mathrm{g}}(w) \right)^2 } \,
        \Bigg]^{\top}\!.
    \end{aligned}
\end{equation}
In our evaluation, we set $W=20$ because the \ac{AC} of both payload and \ac{IAT} decays close to zero within this range.
\\
Third, the generator should preserve dataset-level distributional coverage, capturing the full heavy-tailed diversity of the traffic. Let $\mathcal{P}_{p}^{\mathrm{r}}$ and $\mathcal{P}_{p}^{\mathrm{g}}$ denote the global empirical distributions of the log-transformed payload lengths pooled across all test flows. Let $\mathcal{P}_{\delta}^{\mathrm{r}}$ and $\mathcal{P}_{\delta}^{\mathrm{g}}$ denote the corresponding global distributions for the log-transformed \acp{IAT}. The diversity discrepancy is then defined as a vector containing the \acp{WD} for each feature independently:
\begin{equation}
    \bm{\mathcal{D}}_{\mathrm{div}} = 
    \left[ 
    W_{1}\!\left( \mathcal{P}_{p}^{\mathrm{r}}, \mathcal{P}_{p}^{\mathrm{g}} \right), \quad 
    W_{1}\!\left( \mathcal{P}_{\delta}^{\mathrm{r}}, \mathcal{P}_{\delta}^{\mathrm{g}} \right) 
    \right]^{\top}\!,
\end{equation}
where $W_{1}(\cdot,\cdot)$ denotes the 1D \ac{WD} \cite{arjovsky2017wassersteingan}.
The traffic-generation objective is therefore not a direct closed-form optimization of these discrepancies, but the construction of a probabilistic generator whose parameters are learned by maximum-likelihood surrogates for which $\bm{\mathcal{D}}_{\mathrm{marg}}(\bm{\Omega})$, $\bm{\mathcal{D}}_{\mathrm{temp}}(\bm{\Omega})$, and $\bm{\mathcal{D}}_{\mathrm{div}}(\bm{\Omega})$ are all small on held-out flows. These three quantities define what traffic similarity means in this article and are used later in Section \ref{sec:results} for evaluation.
}

\begin{figure*}[t]
    \centering
    \includegraphics[width=1\linewidth]{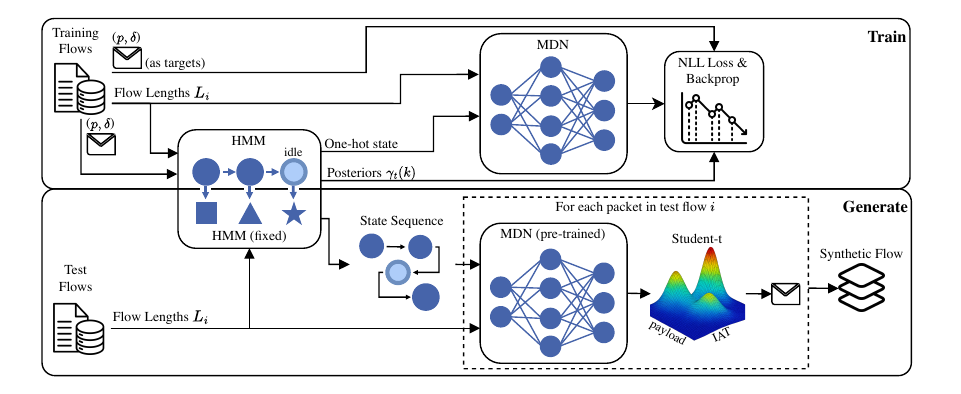}
    \caption{Training and generation pipeline of the proposed traffic generator.}
    \label{fig:system_model}
\end{figure*}

\section{System Model}\label{sec:system_model}
\textcolor{red}{We approximate the real flow distribution with a two stage hybrid generator. The first stage is realized by a factorized latent-state generator that separates state evolution from packet generation within a state. Using \eqref{eq:norm_flow_length} and \eqref{eq:one_hot_state}, let
\begin{equation}\label{eq:conditioning_vec}
\bm{h}_{i,t}^{(k)} = \begin{bmatrix} \left(\bm{e}^{(k)}\right)^\top & \xi_i \end{bmatrix}^{\top} \in \mathbb{R}^{K+1}
\end{equation}
be the conditioning vector for state $k$. The hybrid generator's joint distribution over one normalized flow and one latent-state sequence is
\begin{equation}\label{eq:hybrid_joint}
p_{\bm{\Omega}} \!\left( \bm{z}_{i,1:L_i}, q_{i,1:L_i} \,\middle|\, \xi_i \right) = p_{\bm{\Theta}} \!\left( q_{i,1:L_i} \right) \prod_{t=1}^{L_i} p_{\bm{\varphi}} \!\left( \bm{z}_{i,t} \,\middle|\, \bm{h}_{i,t}^{(q_{i,t})} \right).
\end{equation}
The first factor in \eqref{eq:hybrid_joint} describes the temporal backbone of the flow. Using the initial-state distribution $\bm{\alpha}$ and transition matrix $\bm{A}$, we write
\begin{equation}\label{eq:state_chain}
p_{\bm{\Theta}} \!\left( q_{i,1:L_i} \right) = \alpha_{q_{i,1}} \prod_{t=2}^{L_i} a_{q_{i,t-1},q_{i,t}},
\end{equation}
where $\alpha_k = \Pr(q_{i,1}=k)$ is the initial probability of state $k$ (an element of $\bm{\alpha}$), and $a_{k,\ell} = \Pr(q_{i,t}=\ell \mid q_{i,t-1}=k)$ is the transition probability from state $k$ to state $\ell$ (an element of $\bm{A}$).
The second factor in \eqref{eq:hybrid_joint} describes packet generation conditioned on the state. In our model, this conditional density is represented by the Student-$t$ \ac{MDN} introduced later in this section. We define this as
\begin{equation}\label{eq:hybrid_emission}
p_{\bm{\varphi}} \!\left( \bm{z}_{i,t} \,\middle|\, \bm{h}_{i,t}^{(q_{i,t})} \right) = \sum_{c=1}^{M} \omega_c\!\left(\bm{h}_{i,t}^{(q_{i,t})}; \bm{\varphi}\right) f_c\!\left(\bm{z}_{i,t} \,\middle|\, \bm{h}_{i,t}^{(q_{i,t})}; \bm{\varphi}\right),
\end{equation}
where first element in the sum is the mixture weight of component $c$ parameterized by the network weights $\bm{\varphi}$, and the second element is the corresponding diagonal bivariate Student-$t$ density.}

\textcolor{red}{
Marginalizing the latent-state sequence in \eqref{eq:hybrid_joint} yields the model distribution of one normalized flow:
\begin{equation}
\label{eq:hybrid_marginal}
p_{\bm{\Omega}}
\!\left(
\bm{z}_{i,1:L_i}
\,\middle|\,
\xi_i
\right)
=
\sum_{q_{i,1:L_i}}
p_{\bm{\Omega}}
\!\left(
\bm{z}_{i,1:L_i},
q_{i,1:L_i}
\,\middle|\,
\xi_i
\right).
\end{equation}
\\
This factorization is shown in Fig.~\ref{fig:system_model}. The upper section of the figure illustrates the components and processes involved in the training phase, while the lower section depicts the generation phase that occurs once training is complete. The \ac{HMM} captures when traffic switches between coarse behavioral modes such as buffering, steady streaming, and idle, and how long it remains in each mode, while the \ac{MDN} captures the heavy-tailed and multimodal packet statistics within each mode. In this sense, the hybrid model is a compact surrogate for the objectives. The state process addresses $\mathcal{D}_{\mathrm{temp}}$, and the conditional emission model addresses $\mathcal{D}_{\mathrm{marg}}$ and $\mathcal{D}_{\mathrm{div}}$.}

Standard \ac{HMM} training uses the \ac{EM} algorithm, which alternates between an expectation step (E-step) that computes state posteriors given the current parameters and a maximization step (M-step) that updates the parameters to increase the data likelihood \cite{18626}. Such training often starts from random parameters or from a simple clustering of the data. On heavy-tailed packet traces, this initialization can lead to unstable state distributions and poor local optima. We therefore devote the next subsection to a tailored initialization that already encodes an explicit idle state in the tail, initializes the core states from data driven clusters, and adds weak priors on the transition probabilities. This makes the \ac{HMM} more stable from the first EM iterations and yields state sequences that can reliably supervise the \ac{MDN}.

\subsection{Hidden Markov model: Initialization}\label{subsec:hmm_init}
We first fit a compact \ac{HMM} with single-component diagonal Gaussian emissions to obtain per-packet state assignments along each flow. The goal of this stage is not to be the final generator but to discover coherent states and provide reliable posteriors for the supervised training of the \ac{MDN}. 
The \ac{HMM} parameters are collected in 
\begin{equation}\label{eq:parameter_set}
\bm{\Theta}=\{\bm{\alpha},\bm{A},\{\bm{\mu}_{k},\bm{\Sigma}_{k}\}_{k=1,\ldots,K}\},
\end{equation}
where $\bm{\alpha}\in[0,1]^K$ is the initial-state distribution whose elements add up to one, and $\bm{A}\in[0,1]^{K\times K}$ is the transition matrix where the elements of each row sum to one, $\bm{\mu}_k \in \mathbb{R}^2$ are the per-state means, and $\bm{\Sigma} = \{\bm{\Sigma}_{k}\in\mathbb{R}_{+}^{2}\}$
collects the per-state variance vectors. The diagonal covariance used in the Gaussian kernels is $\operatorname{diag}\!\bigl(\bm{\Sigma}_{k}\bigr)$.

Heavy right tails in \ac{IAT} demand an explicit idle state rather than forcing a single Gaussian to stretch over orders of magnitude. \textcolor{red}{We determine the total number of states $K$ during a pre-processing step prior to initialization.} We therefore start from a core number of states $K_{\mathrm{core}}\in\mathbb{N}$ and create an idle state $K=K_{\mathrm{core}}+1$ when the training traffic exhibits a meaningful tail. 
If the fraction of packets above $\upsilon_{\delta}$ in \eqref{eq:normalized_threshold} on training flows exceeds a small threshold $\theta_{\mathrm{tail}}\in(0,1)$, we activate the idle. Otherwise, we keep $K=K_{\mathrm{core}}$. This rule activates the extra state precisely when the trace contains sufficiently many long gaps to justify a dedicated state, which stabilizes posteriors and reduces the tendency of \ac{EM} to inflate variances in all states to cover extremes.

We initialize the core Gaussian emissions from the joint geometry of payload and timing. Following the standard \ac{HMM} practice, in which state-dependent Gaussians are initialized from a clustering step before EM \cite{18626}, we run $k$-means on the normalized training packets. Specifically, we apply $k$-means with $K_{\mathrm{core}}$ clusters to the pooled set of normalized packets in $\bm{z}$-space obtained by concatenating all training flows. Let $\{\bm{c}_{k}\}_{k=1}^{K_{\mathrm{core}}}\subset\mathbb{R}^{2}$ denote the resulting centroids, and let $\bm{v}_{k}\in\mathbb{R}_{+}^{2}$ denote the componentwise empirical variance of the packets assigned to centroid $\bm{c}_{k}$. We then initialize core state $k$ with mean and diagonal covariance $\bm{\mu}_{k}=\bm{c}_{k}$, and
\begin{equation}
\bm{\Sigma}_{k}=\bm{v}_{k}+\varepsilon_{0}\bm{1},
\end{equation}
where $\bm{1}$ is the length-two vector of ones and $\varepsilon_{0}>0$ is a small variance floor that prevents degenerate covariances.


When the idle state is active, we place its initial mean at a stable, interpretable location. Denote by $\rho_{z}\in\mathbb{R}$ the median of the first coordinate of all training packets in $\bm{z}$-space. The idle mean is then
\begin{equation}
\bm{\mu}_{K}=\begin{bmatrix}\rho_{z}\\ \upsilon_{\delta}\end{bmatrix},\qquad
\bm{\Sigma}_{K}=\begin{bmatrix}\chi\\ 1\end{bmatrix}
\end{equation}
with a broad variance level $\chi>0$ that allows the \ac{EM} step to adapt to the precise spread of long gaps. Initializing the timing coordinate exactly at $\upsilon_{\delta}$ has two effects. First, it creates a clear basin that attracts packets near and above the tail onset, which improves the separation between core and idle states. Second, it prevents the core states from inflating their variances to accommodate rare long pauses, which would blur state boundaries and degrade the quality of the posteriors. Unlike the generic $k$-means initializing above, anchoring a dedicated idle state at a fixed $\upsilon_{\delta}$ is specific to heavy-tailed packet IAT and, to our knowledge, is not part of standard HMM initializations \cite{18626}.

The start distribution is initialized as $\alpha_{k}=1/K$. To stabilize early \ac{EM} iterations we add weak Dirichlet pseudo-counts to the rows of the transition matrix $\bm{A}$, which encourages state continuity but becomes negligible as more data are observed. For the idle state, we use a stronger self pseudo-count and very small pseudo-counts toward core states so that long gaps are captured while transitions back to core states remain possible. The exact prior and its effect on dwell times are discussed in Section~\ref{sec:system_model}--\ref{subsec:hmm_core}.

After each M-step we enforce a minimal variance level, renormalize transition rows, and reset any non-finite parameters. In a standard \ac{HMM}, the latent states are not observed, so the training procedure needs the probability of each state at each time index given the whole sequence. The forward-backward algorithm provides exactly these per-time state probabilities by combining information from the past and the future of the same sequence \cite{18626}. We apply this inference separately on each flow using its own length $L_i$, which avoids learning artificial transitions across flows and yields reliable dwell times. These precautions keep the \ac{HMM} stable and guarantee that the state posteriors are well behaved across packet scales.

Each element of the initialization is tied to a failure mode we observed in real packet traces. The idle state isolates the rare but operationally important long gaps so that core states are dedicated to active behavior. Initializing with $k$-means in the joint space of payload and timing encourages cross-feature coherence and reduces sensitivity to random initializations. Dirichlet priors on transitions encode only weak persistence assumptions. They improve dwell-time realism early in training and are essential when traces are short or dominated by a few bursts. Finally, the variance floor and post-\ac{EM} sanity checks harden the procedure against numerical collapse in heavy tails. Together these choices yield clean state posteriors and a transparent transition structure that the \ac{MDN} can reliably refine in the next stage.

We note that we do not discover the true semantic states or their cardinality from data. The model uses a small state budget $K_{\mathrm{core}}$ that is a tunable hyperparameter, typically ranging from $2$ to $4$. It is set by the practitioner or selected with a light model-selection rule on training flows. The goal is a compact \ac{HMM} that yields stable posteriors and a controllable generator rather than an unsupervised extraction of behavioral modes. 
This choice reflects the \ac{DT} setting, where reproducibility and low overhead retuning are preferable to open-ended state discovery.

\subsection{Hidden Markov model: State process and inference}\label{subsec:hmm_core}

The \ac{HMM} is a $K$-state first order Markov chain that evolves along each flow and selects a state for every packet. We denote the latent state sequence by $q_{i,1:L_i}$ with $q_{i,t}\in\{1,\ldots,K\}$. The chain is parameterized by $\bm{\alpha}$ and $\bm{A}$ with non-negative rows that sum up to one. 
Conditioned on the state $q_{i,t}=k$, the normalized packet $\bm{z}_{i,t}$ is drawn from a diagonal Gaussian with mean $\bm{\mu}_{k}$ and variance vector $\bm{\Sigma}_{k}$.
The joint model for one flow leverages the state chain defined in \eqref{eq:state_chain} and is written as
\begin{equation}
\begin{aligned}
p\!\left(q_{i,1:L_i}, \bm{z}_{i,1:L_i}\right) &= p_{\bm{\Theta}} \!\left( q_{i,1:L_i} \right) \\ &\times \prod_{t=1}^{L_i} \mathcal{N}\!\left( \bm{z}_{i,t};\, \bm{\mu}_{q_{i,t}}, \operatorname{diag}\!\bigl(\bm{\Sigma}_{q_{i,t}}\bigr) \right).
\end{aligned}
\end{equation}
At this stage we restrict the covariances to diagonal form so that the \ac{HMM} remains compact, numerically stable under EM updates, and fast to re-estimate. Cross-feature dependence between payload and \ac{IAT} is mainly handled by the \ac{MDN} later. This division of roles keeps the \ac{HMM} focused on dwell times and state switching without having to stretch Gaussian clouds across heavy right tails.

We estimate $\bm{\Theta}$ in \eqref{eq:parameter_set} on multiple independent flows by concatenating per-flow sequences and supplying a lengths vector $L_i$ so that forward-backward is reset at each flow boundary. It allows the algorithm to sum the expected counts (the estimated number of times states and transitions occurred) across all flows, effectively performing a joint update as per the multi-sequence Baum-Welch algorithm in \cite{18626}, implemented via \cite{hmmlearn}. This enables HMM to estimate the parameters by EM over training flows while respecting the flow boundaries.
The E-step uses the standard forward-backward recursions on each flow $i$ to compute the state posteriors
\begin{equation}\label{eq:state_posteriors}
\gamma_{i,t}^{(k)} = p\!\left(q_{i,t}=k \mid \bm{z}_{i,1:L_i}\right),
\end{equation}
where $\gamma_{i,t}^{(k)} \in [0,1]$ is the posterior probability that packet $(i,t)$ belongs to state $k$, satisfying 
\begin{equation}
\sum_{k=1}^K \gamma_{i,t}^{(k)} = 1,
\end{equation}
and the two-slice marginals
\begin{equation}\label{eq:zeta}
\zeta_{i,t}^{(k,j)} = p\!\left(q_{i,t}=k,\,q_{i,t+1}=j \mid \bm{z}_{i,1:L_i}\right).
\end{equation}
The state posteriors act as weights for the emission update in the M-step.
The M-step starts with the covariance, which we update from weighted second central moments:
\begin{equation}\label{eq:sigma_main}
\bm{\Sigma}_{k} = \frac{\bm{S}_{k}}{R_{k}} + \varepsilon_{0}\bm{1},
\end{equation}
where $\varepsilon_{0}>0$ keeps variances positive in heavy-tail segments. The effective sample size (responsibility mass) using \eqref{eq:state_posteriors} is
\begin{equation}\label{eq:R}
R_{k} = \sum_{i}\sum_{t=1}^{L_i} \gamma_{i,t}^{(k)},
\end{equation}
the weighted sum of squared deviations is
\begin{equation}\label{eq:S}
\bm{S}_{k}
=
\sum_{i}\sum_{t=1}^{L_i}\gamma_{i,t}^{(k)}
\left(\bm{z}_{i,t}-\bm{\mu}_{k}\right)\odot\left(\bm{z}_{i,t}-\bm{\mu}_{k}\right),
\end{equation}
where $\odot$ denotes elementwise product. The state mean uses the same weights as \eqref{eq:R},
\begin{equation}\label{eq:mu}
\bm{\mu}_{k} = \frac{\sum_{i}\sum_{t=1}^{L_i}\gamma_{i,t}^{(k)}\,\bm{z}_{i,t}}{R_{k}}.
\end{equation}

We update the state process from expected start and transition counts, which are accumulated across flows as
\begin{equation}\label{eq:Nstart}
N_k^{\mathrm{start}} = \sum_{i}\gamma_{i,1}^{(k)},
\end{equation}
and
\begin{equation}\label{eq:Nkl}
N_{k,\ell} = \sum_{i}\sum_{t=1}^{L_i-1}\zeta_{i,t}^{(k,\ell)}.
\end{equation}
using \eqref{eq:zeta}. The normalized updates are
\begin{equation}\label{eq:alpha}
\alpha_{k} = \frac{N_k^{\mathrm{start}}}{\sum_{j=1}^{K} N_j^{\mathrm{start}}},
\end{equation}
\begin{equation}\label{eq:A}
a_{k,\ell} = 
\frac{N_{k,\ell} + \Lambda_{k,\ell}}
{\sum_{r=1}^{K}\bigl(N_{k,r} + \Lambda_{k,r}\bigr)}.
\end{equation}
Here $\bm{\Lambda}\in\mathbb{R}_{+}^{K\times K}$ is a Dirichlet prior with
\begin{equation}\label{eq:Lambda}
\Lambda_{k,\ell}=
\begin{cases}
\lambda_{\mathrm{self}}, & k=\ell,\ k\neq K,\\
\lambda_{\mathrm{off}},  & k\neq \ell,\ k\neq K,\\
\lambda_{\mathrm{idle}}, & k=\ell=K,\\
\lambda_{\mathrm{leak}}, & k=K,\ \ell\neq K,
\end{cases}
\end{equation}
where $\bm{\Lambda}$ collects Dirichlet pseudo-counts on transitions, and $\lambda_{\mathrm{self}}, \lambda_{\mathrm{off}}, \lambda_{\mathrm{idle}}, \lambda_{\mathrm{leak}}\in\mathbb{R}_{+}$ are scalar hyperparameters. \textcolor{red}{These values encapsulate prior expectations about the flow dynamics. $\lambda_{\mathrm{self}}$ and $\lambda_{\mathrm{off}}$ direct the persistence and switching among active traffic states, while $\lambda_{\mathrm{idle}}$ and $\lambda_{\mathrm{leak}}$ dictate the tendency to remain in or exit the idle state. Since $N_{k,\ell}$ is an expected transition count, the influence of $\bm{\Lambda}$ decreases as the number of observed packet transitions increases. This is intentional in our offline training, where the prior is used only as a statistical stabilizer that prevents the transition matrix from overfitting to insufficient data.}
The $K$-th row corresponds to the idle state when present.
By construction the update in \eqref{eq:A} is row-normalized, so rows of $\bm{A}$ sum up to one. Flow boundaries are enforced in \eqref{eq:Nkl} by omitting $t=L_i$.
We iterate E- and M-steps with these definitions until the relative improvement in average log-likelihood stalls or a fixed iteration cap is reached. After each M-step we apply the variance floor in \eqref{eq:sigma_main}, replacing any non-finite entries in \eqref{eq:mu} by neutral values to maintain numerical stability, and renormalize rows of $\bm{A}$. The resulting $\bm{\Theta}$ parameterize the next E-step and later drive state path sampling during synthesis.

The initial distribution is set to $\alpha_{k}=1/K$ at the beginning of training and then updated during the training process described. This choice reflects the fact that application flows can begin in any state. The transition prior $\bm{\Lambda}$ retains the diagonal preference from initialization. Core rows use a moderate self-transition pseudo-count $\lambda_{\mathrm{self}}$ and a smaller off-diagonal level $\lambda_{\mathrm{off}}$, which reduces early oscillations without biasing the learned chain once sufficient data accumulate. For the idle state, its row receives a stronger self-prior $\lambda_{\mathrm{idle}}$ and a very small leak level $\lambda_{\mathrm{leak}}$. This mirrors the behavior that once a flow enters a long waiting period it tends to stay there for many packets. In a finite-state HMM, the expected dwell time in state $k$ is
\begin{equation}
    E\left[D_k\right] = 1/\left(1-a_{k,k}\right),
\end{equation}
so increasing the idle state’s $a_{K,K}$ encodes longer idle spans in the simplest possible way \cite{18626}. This design is consistent with Internet traffic measurements, where waiting times are empirically long and often heavy-tailed, which is a key mechanism behind self-similar burstiness at the aggregate level \cite{650143}. At the individual user level inactive and active periods exhibit no characteristic length, allowing arbitrarily long idle or busy spans with non-negligible probability \cite{282603}. We use the same prior across datasets. It acts only as a weak stabilizer during early EM iterations and is quickly dominated by the data. As a result of these initialization modifications, the final transition matrix $\bm{A}$ produces realistic dwell times while remaining data-driven.


We use the soft posteriors $\gamma_{i,t}^{(k)}$ rather than a hard Viterbi path to supervise the \ac{MDN} later. This keeps information about uncertainty in parts of a flow where states switch, for example near the start or end of buffering periods. With soft labels the network can use packets that belong partly to two states instead of treating them as mislabelled. During synthesis we do not replay a single best state sequence from training. We generate new state paths by sampling from the learned $\bm{\alpha}$ and $\bm{A}$. This reintroduces randomness into the state process and produces synthetic flows whose variability is closer to what we see in the test flows.

In summary, we keep the \ac{HMM} compact so that it remains stable and easy to retrain in a digital-twin loop. Diagonal Gaussian emissions with a variance floor and a Dirichlet prior on transitions prevent numerical problems on heavy-tailed \acp{IAT} and short traces. This layer is only used to produce well-behaved state posteriors along each flow, which then supervise the \ac{MDN} in the next subsection.

\subsection{Mixture density network}\label{subsec:mdn}
The \ac{MDN} refines each latent state with a conditional mixture that is expressive enough to reproduce the joint geometry of packet payload and \ac{IAT} and stable enough to train with a small number of hyperparameters. The network conditions on the state of the packet and the length of the flow that the packet is supposed to be located in. Conditioning on the state allows the network to learn the shape of the mixture densities that best represent the typical payload and \ac{IAT} values that are most likely to be observed within that particular state. The choice of including flow length as a conditioning feature comes from our observations in \cite[Fig.~4]{koktas2025state}. We observed that the model struggles to capture \ac{IAT} characteristics of the flows consisting of only one to five packets in the \ac{UDP} dataset. The flow-length (packet-count) feature enables the model to differentiate packet characteristics between long and short flows.


The network maps $\bm{h}_{i,t}^{(k)}$ in \eqref{eq:conditioning_vec} to the parameters of a mixture with $M$ components that models the normalized target $\bm{z}_{i,t}$. In the prior version, we used Gaussian kernels for these components, which worked well on light and medium tails but consistently under-estimated the probability of extreme \acp{IAT} in applications such as Facebook video calls \cite{koktas2025state}. To handle these heavy right tails without changing the overall architecture, we replace each Gaussian component by a Student-$t$ kernel and let the network learn a degrees-of-freedom parameter per component \cite{student_t_ref}.

The \ac{MDN} outputs the parameters of the conditional Student-$t$ mixture in \eqref{eq:hybrid_emission}. We use $M\in\mathbb{N}$ for the number of mixture components and $c\in\{1,\ldots,M\}$ for the component index. Given the conditioning vector $\bm{h}$, it produces mixture weights $\bm{\omega}(\bm{h})\in\mathbb{R}_{+}^{M}$ with
\begin{equation}\label{eq:comp_weight_sum}
\sum_{c=1}^{M}\omega_c(\bm{h})=1,
\end{equation}
together with a component mean $\bm{\kappa}_c(\bm{h})\in\mathbb{R}^{2}$, a positive component scale $\bm{\sigma}_c(\bm{h})\in\mathbb{R}_{+}^{2}$ that acts as a per-coordinate standard deviation, and degrees of freedom $\nu_c(\bm{h})\in\mathbb{R}_{>1}$. Smaller $\nu_c(\bm{h})$ yields heavier tails and assigns non-negligible mass to long \acp{IAT} and occasional payload bursts, whereas larger values approach Gaussian-like behavior. Each mixture component is a diagonal bivariate Student-$t$ in the normalized packet space. The diagonal form keeps the parameter count small and numerically stable, while the mixture captures cross-coordinate dependence without introducing full covariances.
In the implementation, $\bm{\omega}(\bm{h})$ is obtained via a softmax layer to satisfy \eqref{eq:comp_weight_sum}, $\bm{\sigma}_c(\bm{h})$ uses a softplus activation with a small floor, and $\nu_c(\bm{h})$ uses a softplus followed by a constant floor that enforces $\nu_c(\bm{h})>1$. The \ac{MDN} is a feed-forward network with two $\tanh$-activated hidden layers of width $H$, and we keep the same compact configuration across datasets with $M=32$ components, $H=128$ hidden units, and one learned $\nu_c$ per component.

Training follows maximum likelihood with soft supervision from the bootstrap \ac{HMM}. The general formula for the univariate Student-$t$ density with location $\kappa\in\mathbb{R}$, positive scale $\sigma\in\mathbb{R}_{+}$, and degrees of freedom $\nu\in\mathbb{R}_{>1}$ can be written as in \cite[Eq.\,(3)]{student_t_ref}:
\begin{equation}
\label{eq:student_univ}
f_{\nu}\!\left(y \mid \kappa,\sigma\right)
=
c(\nu,\sigma)
\left(
1 + \frac{1}{\nu}
\left(\frac{y-\kappa}{\sigma}\right)^{2}
\right)^{-\frac{\nu+1}{2}},
\end{equation}
where the normalizing constant is
\begin{equation}
c(\nu,\sigma)
=
\frac{\Gamma\bigl((\nu+1)/2\bigr)}
{\Gamma\bigl(\nu/2\bigr)\sqrt{\nu\pi}\,\sigma},
\end{equation}
where $\Gamma(\cdot)$ is the gamma function. Taking natural logarithm gives
\begin{equation}\label{eq:student_log}
\begin{aligned}
\ln f_{\nu}\!\left(y \mid \kappa,\sigma\right)
=&
\log c(\nu,\sigma)
-\frac{\nu+1}{2}\,\\
&\times\log\!\left(
1 + \frac{1}{\nu}
\left(\frac{y-\kappa}{\sigma}\right)^{2}
\right).
\end{aligned}
\end{equation}

For component $c$ and coordinate $d\in\{1,2\}$ the network outputs $\kappa_{c}^{(d)}(\bm{h})$, $\sigma_{c}^{(d)}(\bm{h})$, and $\nu_{c}(\bm{h})$. The corresponding diagonal bivariate Student-$t$ density is the product of its univariate factors,
\begin{equation}
\label{eq:student_diag}
f_{c}\!\left(\bm{z}\mid\bm{h}\right)
=
\prod_{d=1}^{2}
f_{\nu_{c}(\bm{h})}\!\left(z^{(d)} \mid \kappa_{c}^{(d)}(\bm{h}),\sigma_{c}^{(d)}(\bm{h})\right),
\end{equation}
where the same $\nu_{c}(\bm{h})$ is shared across coordinates and each coordinate has its own scale. \textcolor{red}{Using the conditional mixture density given in \eqref{eq:hybrid_emission} the per-sample Student-$t$ \ac{NLL} is
\begin{equation}
\label{eq:mdn_nll_sample}
\ell_{\mathrm{t}}\bigl(\bm{z}_{i,t};\bm{h}_{i,t}^{(k)},\bm{\varphi}\bigr)
=
-\log p_{\bm{\varphi}}\!\left(\bm{z}_{i,t}\mid\bm{h}_{i,t}^{(k)}\right).
\end{equation}}
Expanding \eqref{eq:hybrid_emission} with \eqref{eq:student_diag} and \eqref{eq:student_univ} shows that $\ell_{\mathrm{t}}$ is the negative log of a mixture of diagonal bivariate Student-$t$ components, with the constant and $\log\!\left(1+(\cdot)^{2}/\nu_{c}\right)$ terms coming from \eqref{eq:student_log} and accumulated across $d$ and $c$.

The MDN is trained on mini-batches $\mathcal{B}$ drawn from the training packets. For each $(i,t)\in\mathcal{B}$ we plug the observed $\bm{z}_{i,t}$ into \eqref{eq:mdn_nll_sample} and minimize the weighted objective
\begin{equation}
\label{eq:mdn_weighted_loss}
\mathcal{L}(\bm{\varphi})
=
\sum_{(i,t)\in\mathcal{B}}
\beta_{i,t}\,
\ell_{\mathrm{t}}\bigl(\bm{z}_{i,t};\bm{h}_{i,t}^{(k)},\bm{\varphi}\bigr),
\end{equation}
where $\beta_{i,t}$ are packet weights defined by the \ac{HMM} posteriors and balanced across states in the next paragraph. 
This process is equivalent to standard maximum-likelihood training of a conditional density model. The network sees packets from the dataset together with their state and flow length features, and adjusts $\bm{\varphi}$ so that
the density in \eqref{eq:hybrid_emission} assigns high probability to the training packets. Sampling from the mixture in \eqref{eq:hybrid_emission} happens only in the synthesis stage described later.

The MDN dataset is formed by concatenating per-state slices. For each state $k$ we keep packet $(i,t)$ if the posterior $\gamma_{i,t}^{(k)}$ from the bootstrap \ac{HMM} exceeds a small threshold. The idle state uses a lower threshold so that rare long idle packets are not discarded.
For every retained packet we create a target $\bm{z}_{i,t}$, an input $\bm{h}_{i,t}^{(k)}$, and a weight $\beta_{i,t}=\gamma_{i,t}^{(k)}$. Inside each state these posteriors act as soft labels, so packets with higher state probability have proportionally more influence on the loss. To prevent very frequent states from dominating the objective, we multiply all weights of a given state by a constant so that the total weight per state stays on a comparable scale. 

During synthesis, the \ac{HMM} provides a state sequence for each test flow. For packet $t$ in state $k$, the MDN evaluates $\bm{\omega},\bm{\varphi}$ at $\bm{h}_{i,t}^{(k)}$, draws a component index from $\bm{\omega}$ and samples a diagonal bivariate Student-$t$. The result is mapped back to the original domain by inverting the normalization and the log transforms and clip payload and IAT to the observed operational ranges (\ac{MTU} and a practical timeout). An optional temperature on the component sampling for the idle state makes extremely long gaps marginally more likely.  


\begin{table}[t]
\centering
\caption{Comparison of temporal, diversity, and marginal discrepancy metrics of different models on various datasets.}
\label{tab:comp_temp_div}
\resizebox{\columnwidth}{!}{%
\begin{tabular}{llllllll}
\hline
                                             &                 & \multicolumn{2}{c}{$\mathcal{D}_{\mathrm{marg}}$} & \multicolumn{2}{c}{$\mathcal{D}_{\mathrm{temp}}$ (\ac{AC})} & \multicolumn{2}{c}{$\mathcal{D}_{\mathrm{div}}$ (\ac{WD})} \\
Model                                        & Dataset         & Payload                 & IAT                     & Payload                      & IAT                          & Payload                      & IAT                         \\ \hline
\multirow{4}{*}{Proposed}                    & Google Home UDP & $\mathbf{1.510}$        & $\mathbf{0.869}$        & $0.101$                      & \underline{$0.070$}          & $\mathbf{0.151}$             & \underline{$0.071$}         \\
                                             & Facebook Audio  & \underline{$1.478$}     & \underline{$0.533$}        & \underline{$0.066$}          & $\mathbf{0.054}$             & $0.749$                      & \underline{$0.027$}         \\
                                             & Facebook Video  & $\mathbf{1.736}$        & $0.674$                 & $0.136$                      & $\mathbf{0.119}$          & $\mathbf{0.698}$             & $0.476$                     \\
                                             & FTPS            & $\mathbf{1.769}$        & $\mathbf{0.533}$        & \underline{$0.068$}          & \underline{$0.046$}          & \underline{$0.013$}          & $\mathbf{0.001}$            \\ \hline
\multirow{4}{*}{\cite{9320384}}              & Google Home UDP & $1.536$                 & \underline{$1.111$}     & $0.177$                      & $0.133$                      & $0.384$                      & $0.123$                     \\
                                             & Facebook Audio  & $1.687$                 & $1.893$                 & $0.323$                      & $0.264$                      & $1.096$                      & $0.946$                     \\
                                             & Facebook Video  & \underline{$1.760$}     & $1.582$                 & $0.263$                      & $0.427$                      & \underline{$0.815$}          & $0.682$                     \\
                                             & FTPS            & $1.924$                 & $1.413$                 & $0.270$                      & $0.283$                      & $2.112$                      & $1.091$                     \\ \hline
\multirow{4}{*}{\cite{10000731}}             & Google Home UDP & $1.929$                 & $1.206$                 & $0.123$                      & $0.077$                      & $0.542$                      & $0.648$                     \\
                                             & Facebook Audio  & $1.901$                 & $1.990$                 & $0.115$                      & $0.070$                      & \underline{$0.733$}          & $\mathbf{0.024}$            \\
                                             & Facebook Video  & $1.999$                 & $\mathbf{0.605}$        & $0.126$                      & $0.173$                      & $3.263$                      & $0.506$                     \\
                                             & FTPS            & $1.919$                 & \underline{$0.538$}     & $0.214$                      & $\mathbf{0.041}$             & $\mathbf{0.005}$             & $\mathbf{0.001}$            \\ \hline
\multirow{4}{*}{\cite{qu2024trafficgpt}}     & Google Home UDP & $1.635$                 & $1.962$                 & $\mathbf{0.081}$             & $0.128$                      & \underline{$0.160$}          & $\mathbf{0.060}$            \\
                                             & Facebook Audio  & $\mathbf{1.366}$        & $1.986$                 & $0.080$                      & \underline{$0.055$}          & $1.830$                      & $0.240$                     \\
                                             & Facebook Video  & $2.000$                 & $1.997$                 & $0.141$                      & $0.161$                      & $0.930$                      & $\mathbf{0.390}$            \\
                                             & FTPS            & \underline{$1.829$}     & $1.939$                 & $0.161$                      & $0.224$                      & $1.642$                      & $0.288$                     \\ \hline
\multirow{4}{*}{\cite{9145646}}              & Google Home UDP & $1.958$                 & $1.991$                 & \underline{$0.100$}          & $\mathbf{0.058}$             & $0.151$                      & $0.129$                     \\
                                             & Facebook Audio  & $1.973$                 & $0.892$                 & $0.145$                      & $0.187$                      & $0.778$                      & $0.177$                     \\
                                             & Facebook Video  & $2.000$                 & \underline{$0.617$}     & \underline{$0.105$}          & \underline{$0.129$}                      & $2.845$                      & \underline{$0.470$}         \\
                                             & FTPS            & $1.923$                 & $\mathbf{0.533}$        & $\mathbf{0.061}$             & \underline{$0.174$}          & $0.086$          & $\mathbf{0.001}$            \\ \hline
\multirow{4}{*}{\cite{NEURIPS2019_c9efe5f2}} & Google Home UDP & \underline{$1.534$}     & $1.518$                 & $0.101$                      & $0.130$                      & $0.463$                      & $0.240$                     \\
                                             & Facebook Audio  & $1.978$                 & $\mathbf{0.409}$        & $\mathbf{0.063}$             & $0.359$                      & $\mathbf{0.441}$             & $0.076$                     \\
                                             & Facebook Video  & $1.926$                 & $0.771$                 & $\mathbf{0.101}$             & $0.130$                      & $1.880$                      & $0.485$                     \\
                                             & FTPS            & $1.945$                 & $0.541$                 & $0.200$                      & $0.637$                      & $1.952$                      & $0.004$                     \\ \hline
\end{tabular}
}
\end{table}

In summary, the Student-t mixture lets the MDN capture rare but important extremes without making sampling complicated. Using diagonal components in a mixture gives a stable way to model joint shapes with a moderate number of parameters. The state encoding tells the network which traffic state it is in, and the flow-length input passes session-scale effects down to individual packets. We train the MDN with weights derived from the HMM posteriors so that it learns from the state structure without being dominated by a single state. Together, these choices make the MDN both robust in practice and easy to interpret.

\section{Results} \label{sec:results}
The evaluation is structured around the three core discrepancy metrics formalized in Section~\ref{subsec:formal_problem}: marginal fidelity $\bm{\mathcal{D}}_{\mathrm{marg}}$, short-range temporal correlation $\bm{\mathcal{D}}_{\mathrm{temp}}$, and diversity $\bm{\mathcal{D}}_{\mathrm{div}}$. We compare our proposed hybrid generator against the state-of-the-art neural, transformer, and statistical baselines discussed in Section~\ref{sec:Introduction}-\ref{subsec:related_work}, which we implemented to the best of our knowledge. To ensure a fair and rigorous comparison, all models are evaluated on the four datasets introduced in Section~\ref{sec:dataset_notation} using identical pre-processing, post-processing, and flow-wise train/test splits.

\textcolor{red}{The quantitative results across all three objectives are presented in Table~\ref{tab:comp_temp_div}, where boldface entries indicate the best result and underlined entries indicate the second-best within each dataset and metric. 
As established, evaluating $\bm{\mathcal{D}}_{\mathrm{temp}}$ and $\bm{\mathcal{D}}_{\mathrm{div}}$ on log-transformed data ensures the metrics accurately reflect the structural rhythm and global heavy-tailed spread without being exclusively dominated by a few massive outlier packets. Complementing this, $\bm{\mathcal{D}}_{\mathrm{marg}}$ penalizes models that fail to capture exact local modes, such as strict \ac{MTU} payload boundaries or instantaneous hardware-level bursts.}

\textcolor{red}{The results demonstrate that the proposed generator consistently excels at $\bm{\mathcal{D}}_{\mathrm{marg}}$, achieving the best or second-best scores across nearly all datasets. This confirms its superior ability to accurately reproduce local traffic modes without suffering from the severe mode collapse seen in continuous baselines. Simultaneously, it maintains competitive $\bm{\mathcal{D}}_{\mathrm{div}}$ and $\bm{\mathcal{D}}_{\mathrm{temp}}$, proving it successfully captures both the micro-structure and the macroscopic heavy-tailed spread of the real traces.}

\textcolor{red}{Notably, while the discrete baseline \cite{9145646} often achieves low $\bm{\mathcal{D}}_{\mathrm{temp}}$, this is largely an artifact of its design. By forcing continuous data into fixed bins, it artificially smooths out extreme variations. This quantization yields excellent \ac{AC} scores but sacrifices the true continuous distribution, leading to poor performance on both the fidelity and diversity metrics. In contrast, our proposed model closely tracks these temporal scores while successfully preserving the exactness and continuous heavy-tailed nature of the traffic.}
\textcolor{red}{To directly compare the efficacy of attention mechanisms against our statistical approach, we implemented a compact decoder-only transformer adapted from \cite{qu2024trafficgpt}, trained on quantized payload and \ac{IAT} tokens. Since the transformer treats traffic generation as a discrete grammar problem, it struggles to model the continuous, heavy-tailed nature of \acp{IAT}, resulting in high $\bm{\mathcal{D}}_{\mathrm{marg}}$ errors for timing across four datasets.}

\begin{table}[t]
    \centering
    \caption{\textcolor{red}{Memory footprint and inference time}}
    \label{tab:footprint_inference}
    \resizebox{\columnwidth}{!}{%
        \begin{tabular}{lccc}
            \hline
            Model                     & Parameters    & Footprint (MB) & Time (s)\textsuperscript{a} \\ \hline
            \cite{NEURIPS2019_c9efe5f2} & 67,275        & 0.2566         & 0.0429                      \\
            \cite{9320384}              & $\sim$312,000 & 1.1941         & 0.1099                      \\
            \cite{9145646}              & N/A           & 0.0134         & 0.1298                      \\
            Proposed (no reject)      & $\sim$42,000  & 0.1640         & 0.1851                      \\
            Proposed                  & $\sim$42,000  & 0.1640         & 0.2348                      \\
            \cite{10000731}             & $\sim$480     & 0.0537         & 3.2884                      \\
            \cite{qu2024trafficgpt}     & 885,248       & 4.3770         & 44.9310                     \\ \hline
            \multicolumn{4}{l}{\rule{0pt}{3ex}\textsuperscript{a}\footnotesize Average time to generate 5,000 packets using a single-threaded CPU.}
        \end{tabular}%
    }
\end{table}

\textcolor{red}{In terms of computational complexity, the proposed generator scales linearly with the total number of packets $N_{\mathrm{pkt}} = \sum_i L_i$. The \ac{HMM} parameter estimation via the Baum-Welch algorithm has a time complexity of $\mathcal{O}(K^2 N_{\mathrm{pkt}})$, while the \ac{MDN} training scales as $\mathcal{O}(N_{\mathrm{pkt}} H M)$. Practically, on the $5$~MB Facebook video trace, the complete training and generation process takes approximately $7$ minutes. A portion of this runtime is due to the constant overhead of evaluating log and gamma functions for the Student-$t$ negative log-likelihood in \eqref{eq:mdn_nll_sample}. Further optimization could reduce runtime, which we leave for future work.}

\begin{figure}[t]
    \centering

    \begin{subfigure}[t]{\columnwidth}
        \centering
        \begin{tikzpicture}
            \begin{axis}[xmode=log, filter discard warning=false, unbounded coords=discard,
                width=\cdfaxiswidth,
                height=0.5\linewidth,
                scale only axis,    
                grid style={line width=.2pt, gray!50},
                major grid style={line width=.3pt, gray!60},
                minor tick num=1,
                tick label style={font=\small},
                ylabel style={font=\small},
                xlabel style={font=\small},
                xlabel={},
                ylabel={CDF},
                xmin=1, xmax=1500,
                ymin=0, ymax=1.05,
                grid=both,
                legend cell align={left},
                legend pos=north west,
                legend style={font=\footnotesize, draw=black, fill=white, fill opacity=.8},
            ]
            \addplot[kit-purple100, solid, thick]
                table [x index=0, y index=1, col sep=comma]
                {data/payload_cdf_log_algo_v3_3_UDP.csv};
            \addlegendentry{Real}

            \addplot[black, loosely dashed, thick]
                table [x index=0, y index=2, col sep=comma]
                {data/payload_cdf_log_gen_iot_v2_UDP.csv};
            \addlegendentry{\cite{9320384}}

            \addplot[black, dashdotted, thick]
                table [x index=0, y index=2, col sep=comma]
                {data/payload_cdf_log_necstgen_UDP.csv};
            \addlegendentry{\cite{10000731}}

            \addplot[black, dash pattern=on 8pt off 3pt, thick]
                table [x index=0, y index=2, col sep=comma]
                {data/payload_cdf_log_trafficgpt10_UDP.csv};
            \addlegendentry{\cite{qu2024trafficgpt} (Adapted)}

            \addplot[black, densely dotted, thick]
                table [x index=0, y index=2, col sep=comma]
                {data/payload_cdf_log_icetran_v2_UDP.csv};
            \addlegendentry{\cite{9145646}}

            \addplot[black, dashdotdotted, thick]
                table [x index=0, y index=2, col sep=comma]
                {data/payload_cdf_log_timegan_v2_UDP.csv};
            \addlegendentry{\cite{NEURIPS2019_c9efe5f2}}

            \addplot[kit-cyan100, densely dashed, very thick]
                table [x index=0, y index=2, col sep=comma]
                {data/payload_cdf_log_algo_v3_3_UDP.csv};
            \addlegendentry{Proposed}
            \end{axis}
        \end{tikzpicture}
        \caption{Payload length (bytes, log scaled)}
        \label{fig:cdf_payload_comp_udp}
    \end{subfigure}
    \hfill
    \begin{subfigure}[t]{\columnwidth}
        \centering
        \begin{tikzpicture}
            \begin{axis}[xmode=log, filter discard warning=false, unbounded coords=discard,
                width=\cdfaxiswidth,
                height=0.4\linewidth,
                scale only axis,
                grid style={line width=.2pt, gray!50},
                major grid style={line width=.3pt, gray!60},
                minor tick num=1,
                scaled x ticks=false,
                xticklabel style={
                    /pgf/number format/fixed,
                    /pgf/number format/precision=3
                },
                tick label style={font=\small},
                ylabel style={font=\small},
                xlabel style={font=\small},
                xlabel={},
                ylabel={CDF},
                xmin=0.005,
                ymin=0, ymax=1.02,
                grid=both,
                legend cell align={left},
                legend pos=south east,
                legend style={font=\footnotesize, draw=black, fill=white, fill opacity=.8},
            ]
            \addplot[kit-purple100, solid, thick]
                table [x index=0, y index=1, col sep=comma]
                {data/time_cdf_log_algo_v3_3_UDP.csv};
            \addlegendentry{Real}

            \addplot[black, loosely dashed, thick]
                table [x index=0, y index=2, col sep=comma]
                {data/time_cdf_log_gen_iot_v2_UDP.csv};
            \addlegendentry{\cite{9320384}}

            \addplot[black, dashdotted, thick]
                table [x index=0, y index=2, col sep=comma]
                {data/time_cdf_log_necstgen_UDP.csv};
            \addlegendentry{\cite{10000731}}

            \addplot[black, dash pattern=on 8pt off 3pt, thick]
                table [x index=0, y index=2, col sep=comma]
                {data/time_cdf_log_trafficgpt10_UDP.csv};
            \addlegendentry{\cite{qu2024trafficgpt} (Adapted)}

            \addplot[black, densely dotted, thick]
                table [x index=0, y index=2, col sep=comma]
                {data/time_cdf_log_icetran_v2_UDP.csv};
            \addlegendentry{\cite{9145646}}

            \addplot[black, dashdotdotted, thick]
                table [x index=0, y index=2, col sep=comma]
                {data/time_cdf_log_timegan_v2_UDP.csv};
            \addlegendentry{\cite{NEURIPS2019_c9efe5f2}}

            \addplot[kit-cyan100, densely dashed, very thick]
                table [x index=0, y index=2, col sep=comma]
                {data/time_cdf_log_algo_v3_3_UDP.csv};
            \addlegendentry{Proposed}
            \end{axis}
        \end{tikzpicture}
        \caption{\ac{IAT} (seconds, log scaled)}
        \label{fig:cdf_iat_comp_udp}
    \end{subfigure}

    \caption{Average per-flow \ac{CDF} comparison of payload length and \ac{IAT} for \textcolor{red}{Google Home} \ac{UDP} traffic.}
    \label{fig:cdf_payload_iat_UDP}
\end{figure}

\textcolor{red}{To put these operational costs into perspective, Table~\ref{tab:footprint_inference} evaluates the memory footprint and inference time of all models trained on the Facebook video dataset. To assess algorithmic complexity directly, evaluations were restricted to a single-threaded \ac{CPU} generating exactly $5{,}000$ synthetic packets. The results illustrate varying architectural tradeoffs. \cite{qu2024trafficgpt} faces significant computational overhead on a \ac{CPU} due to its autoregressive nature. Conversely, models like \cite{NEURIPS2019_c9efe5f2} and \cite{9320384} achieve high throughput by decoding data batches simultaneously. However, their reliance on fixed-size matrices restricts the dynamic generation of arbitrary flow lengths. \cite{10000731} requires minimal memory but exhibits higher inference latency due to the execution overhead of its step-by-step sliding window loop. The purely statistical discrete model \cite{9145646} minimizes both footprint and generation time, but as established, this quantization inherently sacrifices the fidelity of extreme heavy-tailed \acp{IAT}.}

\begin{figure}[t]
    \centering

    \begin{subfigure}[t]{\columnwidth}
        \centering
        \begin{tikzpicture}
            \begin{axis}[xmode=log, filter discard warning=false, unbounded coords=discard,
                width=\cdfaxiswidth,
                height=0.5\linewidth,
                scale only axis,   
                grid style={line width=.2pt, gray!50},
                major grid style={line width=.3pt, gray!60},
                minor tick num=1,
                tick label style={font=\small},
                ylabel style={font=\small},
                xlabel style={font=\small},
                xlabel={},
                ylabel={CDF},
                xmin=1, xmax=5000,
                ymin=0, ymax=1.05,
                grid=both,
                legend cell align={left},
                legend pos=south east,
                legend style={font=\footnotesize, draw=black, fill=white, fill opacity=.8},
            ]
            \addplot[kit-purple100, solid, thick]
                table [x index=0, y index=1, col sep=comma]
                {data/payload_cdf_log_algo_v3_3_facebook_audio2a.csv};
            \addlegendentry{Real}

            \addplot[black, dashdotted, thick]
                table [x index=0, y index=2, col sep=comma]
                {data/payload_cdf_log_gen_iot_v2_facebook_audio2a.csv};
            \addlegendentry{\cite{9320384}}

            \addplot[kit-orange100, loosely dashed, thick]
                table [x index=0, y index=2, col sep=comma]
                {data/payload_cdf_log_necstgen_v3_facebook_audio2a.csv};
            \addlegendentry{\cite{10000731}}

            \addplot[black, dash pattern=on 8pt off 3pt, thick]
                table [x index=0, y index=2, col sep=comma]
                {data/payload_cdf_log_trafficgpt10_facebook_audio2a.csv};
            \addlegendentry{\cite{qu2024trafficgpt} (Adapted)}

            \addplot[black, densely dotted, thick]
                table [x index=0, y index=2, col sep=comma]
                {data/payload_cdf_log_icetran_v2_facebook_audio2a.csv};
            \addlegendentry{\cite{9145646}}

            \addplot[kit-green100, dashdotdotted, thick]
                table [x index=0, y index=2, col sep=comma]
                {data/payload_cdf_log_timegan_v2_facebook_audio2a.csv};
            \addlegendentry{\cite{NEURIPS2019_c9efe5f2}}

            \addplot[kit-cyan100, densely dashed, very thick]
                table [x index=0, y index=2, col sep=comma]
                {data/payload_cdf_log_algo_v3_3_facebook_audio2a_2.csv};
            \addlegendentry{Proposed}
            \end{axis}
        \end{tikzpicture}
        \caption{Payload length (bytes, log scaled)}
        \label{fig:cdf_payload_comp_facebook_audio}
    \end{subfigure}
    \hfill
    \begin{subfigure}[t]{\columnwidth}
        \centering
        \begin{tikzpicture}
            \begin{axis}[xmode=log, filter discard warning=false, unbounded coords=discard,
                width=\cdfaxiswidth,
                height=0.4\linewidth,
                scale only axis,
                grid style={line width=.2pt, gray!50},
                major grid style={line width=.3pt, gray!60},
                minor tick num=1,
                scaled x ticks=false,
                xticklabel style={
                    /pgf/number format/fixed,
                    /pgf/number format/precision=3
                },
                tick label style={font=\small},
                ylabel style={font=\small},
                xlabel style={font=\small},
                xlabel={},
                ylabel={CDF},
                xmin=0.006,
                ymin=0.0, ymax=1.01,
                grid=both,
                legend cell align={left},
                legend pos=south east,
                legend style={font=\footnotesize, draw=black, fill=white, fill opacity=.8},
            ]
            \addplot[kit-purple100, solid, thick]
                table [x index=0, y index=1, col sep=comma]
                {data/time_cdf_log_algo_v3_3_facebook_audio2a.csv};
            \addlegendentry{Real}

            \addplot[black, dashdotted, thick]
                table [x index=0, y index=2, col sep=comma]
                {data/time_cdf_log_gen_iot_v2_facebook_audio2a.csv};
            \addlegendentry{\cite{9320384}}

            \addplot[kit-orange100, loosely dashed, thick]
                table [x index=0, y index=2, col sep=comma]
                {data/time_cdf_log_necstgen_v3_facebook_audio2a.csv};
            \addlegendentry{\cite{10000731}}

            \addplot[black, dash pattern=on 8pt off 3pt, thick]
                table [x index=0, y index=2, col sep=comma]
                {data/time_cdf_log_trafficgpt10_facebook_audio2a.csv};
            \addlegendentry{\cite{qu2024trafficgpt} (Adapted)}

            \addplot[black, densely dotted, thick]
                table [x index=0, y index=2, col sep=comma]
                {data/time_cdf_log_icetran_v2_facebook_audio2a.csv};
            \addlegendentry{\cite{9145646}}

            \addplot[kit-green100, dashdotdotted, thick]
                table [x index=0, y index=2, col sep=comma]
                {data/time_cdf_log_timegan_v2_facebook_audio2a.csv};
            \addlegendentry{\cite{NEURIPS2019_c9efe5f2}}

            \addplot[kit-cyan100, densely dashed, very thick]
                table [x index=0, y index=2, col sep=comma]
                {data/time_cdf_log_algo_v3_3_facebook_audio2a_2.csv};
            \addlegendentry{Proposed}
            \end{axis}
        \end{tikzpicture}
        \caption{\ac{IAT} (seconds, log scaled)}
        \label{fig:cdf_iat_comp_facebook_audio}
    \end{subfigure}

    \caption{Average per-flow \ac{CDF} comparison of payload length and \ac{IAT} for Facebook audio traffic.}
    \label{fig:cdf_payload_iat_facebook_audio}
\end{figure}

\textcolor{red}{The proposed hybrid generator strikes an optimal balance for deployment. It maintains a parameter footprint of just $0.164$~MB. This is more than twenty times smaller than the $4.38$~MB footprint of the transformer baseline. During inference, it operates at speeds closely comparable to the discrete and batched generators. To isolate the computational cost of our architectural components, an ablation study was conducted by temporarily disabling the rejection sampling loop during inference. Without the resampling of operationally invalid packets, the raw inference time decreased from $0.2348$~s to $0.1851$~s, demonstrating the high efficiency of the core hybrid architecture. While massive centralized models remain attractive when large compute clusters and vast datasets (e.g., $190$ gigabytes as in \cite{qu2024trafficgpt}) are available, our compact model is uniquely suited for local \ac{RAN} \acp{DT}, where low memory pressure, real-time generation, and fast retraining under dynamic local traffic conditions are critical.}

\textcolor{red}{To complement the introduced metrics with a new perspective, we provide a visual analysis of the average per-flow \acp{CDF} in Figs.~\ref{fig:cdf_payload_iat_UDP}, \ref{fig:cdf_payload_iat_facebook_audio}, \ref{fig:cdf_payload_iat_facebook_video}, and \ref{fig:cdf_payload_iat_ftps}. Unlike the global \ac{WD} evaluation, which pools all packets and can be dominated by a few massive flows, the average per-flow \ac{CDF} assigns equal weight to every flow. This perspective allows that the structural diversity of short, interactive flows is evaluated on equal weight with long, high volume flows.}

\textcolor{red}{As observed in Figs.~\ref{fig:cdf_payload_iat_UDP} and \ref{fig:cdf_payload_iat_facebook_audio}, the real payload distributions exhibit highly discrete, step-like structures. These arise naturally from application-layer messaging patterns and network-layer \ac{MTU} limits. Continuous generative models like \cite{NEURIPS2019_c9efe5f2} and \cite{10000731} struggle here; they attempt to fit smooth continuous curves to the data, which blurs these sharp steps. Conversely, discrete baselines like \cite{9145646} capture the step-like nature by grouping data into fixed bins, but this artificial rounding causes them to miss exact packet sizes. The transformer \cite{qu2024trafficgpt} excels at matching these discrete steps, visibly achieving the highest payload fidelity on the Facebook Audio trace. The proposed model is fundamentally continuous, but the \ac{MDN} learns narrow, highly concentrated peaks around the true packet sizes. After rounding the output to the nearest byte, it closely tracks the discrete steps and remains competitive, offering a practical middle ground without the massive parameter overhead of a transformer.}

\begin{figure}[t]
    \centering

    \begin{subfigure}[t]{\columnwidth}
        \centering
        \begin{tikzpicture}
            \begin{axis}[xmode=log, filter discard warning=false, unbounded coords=discard,
                width=\cdfaxiswidth, 
                height=0.5\linewidth,    
                scale only axis,
                grid style={line width=.2pt, gray!50},
                major grid style={line width=.3pt, gray!60},
                minor tick num=1,
                tick label style={font=\small},
                ylabel style={font=\small},
                xlabel style={font=\small},
                xlabel={},
                ylabel={CDF},
                xmin=1, xmax=2000,
                ymin=0, ymax=1.05,
                grid=both,
                legend cell align={left},
                legend pos=north west,
                legend style={font=\footnotesize, draw=black, fill=white, fill opacity=.8},
            ]
            \addplot[kit-purple100, solid, thick]
                table [x index=0, y index=1, col sep=comma]
                {data/payload_cdf_algo_v3_3_facebook_video1a.csv};
            \addlegendentry{Real}


            \addplot[black, loosely dashed, thick]
                table [x index=0, y index=2, col sep=comma]
                {data/payload_cdf_log_gen_iot_v2_facebook_video1a.csv};
            \addlegendentry{\cite{9320384}}

            \addplot[black, dash pattern=on 5pt off 2pt on 2pt off 2pt, thick]
                table [x index=0, y index=2, col sep=comma]
                {data/payload_cdf_log_necstgen_v3_facebook_video1a.csv};
            \addlegendentry{\cite{10000731}}

            \addplot[black, dash pattern=on 8pt off 3pt, thick]
                table [x index=0, y index=2, col sep=comma]
                {data/payload_cdf_log_trafficgpt10_facebook_video1a.csv};
            \addlegendentry{\cite{qu2024trafficgpt} (Adapted)}

            \addplot[black, densely dotted, thick]
                table [x index=0, y index=2, col sep=comma]
                {data/payload_cdf_log_icetran_v2_facebook_video1a.csv};
            \addlegendentry{\cite{9145646}}

            \addplot[black, dashdotdotted, thick]
                table [x index=0, y index=2, col sep=comma]
                {data/payload_cdf_log_timegan_v2_facebook_video1a.csv};
            \addlegendentry{\cite{NEURIPS2019_c9efe5f2}}


            \addplot[kit-cyan100, densely dashed, very thick]
                table [x index=0, y index=2, col sep=comma]
                {data/payload_cdf_log_algo_v3_3_facebook_video1a.csv};
            \addlegendentry{Proposed}
            \end{axis}
        \end{tikzpicture}
        \caption{Payload length (bytes, log scaled)}
        \label{fig:cdf_payload_comp_facebook_video}
    \end{subfigure}
    \hfill
    \begin{subfigure}[t]{\columnwidth}
        \centering
        \begin{tikzpicture}
            \begin{axis}[xmode=log, filter discard warning=false, unbounded coords=discard,
                width=\cdfaxiswidth,
                height=0.4\linewidth,
                scale only axis,
                grid style={line width=.2pt, gray!50},
                major grid style={line width=.3pt, gray!60},
                minor tick num=1,
                scaled x ticks=false,
                xticklabel style={
                    /pgf/number format/fixed,
                    /pgf/number format/precision=3
                },
                tick label style={font=\small},
                ylabel style={font=\small},
                xlabel style={font=\small},
                xlabel={},
                ylabel={CDF},
                xmin=0.0055,
                ymin=0.0, ymax=1.005,
                grid=both,
                legend cell align={left},
                legend pos=south east,
                legend style={font=\footnotesize, draw=black, fill=white, fill opacity=.8},
            ]
            \addplot[kit-purple100, solid, thick]
                table [x index=0, y index=1, col sep=comma]
                {data/time_cdf_log_algo_v3_3_facebook_video1a.csv};
            \addlegendentry{Real}


            \addplot[black, loosely dashed, thick]
                table [x index=0, y index=2, col sep=comma]
                {data/time_cdf_log_gen_iot_v2_facebook_video1a.csv};
            \addlegendentry{\cite{9320384}}

            \addplot[kit-orange100, dash pattern=on 5pt off 2pt on 2pt off 2pt, thick]
                table [x index=0, y index=2, col sep=comma]
                {data/time_cdf_log_necstgen_v3_facebook_video1a.csv};
            \addlegendentry{\cite{10000731}}

            \addplot[black, dash pattern=on 8pt off 3pt, thick]
                table [x index=0, y index=2, col sep=comma]
                {data/time_cdf_log_trafficgpt10_facebook_video1a.csv};
            \addlegendentry{\cite{qu2024trafficgpt} (Adapted)}

            \addplot[black, densely dotted, thick]
                table [x index=0, y index=2, col sep=comma]
                {data/time_cdf_log_icetran_v2_facebook_video1a.csv};
            \addlegendentry{\cite{9145646}}

            \addplot[kit-green100, dashdotdotted, thick]
                table [x index=0, y index=2, col sep=comma]
                {data/time_cdf_log_timegan_v2_facebook_video1a.csv};
            \addlegendentry{\cite{NEURIPS2019_c9efe5f2}}


            \addplot[kit-cyan100, densely dashed, very thick]
                table [x index=0, y index=2, col sep=comma]
                {data/time_cdf_log_algo_v3_3_facebook_video1a.csv};
            \addlegendentry{Proposed}
            \end{axis}
        \end{tikzpicture}
        \caption{\ac{IAT} (seconds, log scaled)}
        \label{fig:cdf_iat_comp_facebook_video}
    \end{subfigure}

    \caption{Average per-flow \ac{CDF} comparison of payload length and \ac{IAT} for Facebook video traffic.}
    \label{fig:cdf_payload_iat_facebook_video}
\end{figure}

\begin{figure}[t]
    \centering

    \begin{subfigure}[t]{\columnwidth}
        \centering
        \begin{tikzpicture}
            \begin{axis}[xmode=log, filter discard warning=false, unbounded coords=discard,
                width=\cdfaxiswidth,
                height=0.5\linewidth,
                scale only axis,   
                grid style={line width=.2pt, gray!50},
                major grid style={line width=.3pt, gray!60},
                minor tick num=1,
                tick label style={font=\small},
                ylabel style={font=\small},
                xlabel style={font=\small},
                xlabel={},
                ylabel={CDF},
                xmin=1, xmax=2000,
                ymin=0, ymax=1.05,
                grid=both,
                legend cell align={left},
                legend pos=north west,
                legend style={font=\footnotesize, draw=black, fill=white, fill opacity=.8},
            ]
            \addplot[kit-purple100, solid, thick]
                table [x index=0, y index=1, col sep=comma]
                {data/payload_cdf_log_algo_v3_3_ftps_down_1a.csv};
            \addlegendentry{Real}

            \addplot[black, dashdotted, thick]
                table [x index=0, y index=2, col sep=comma]
                {data/payload_cdf_log_gen_iot_v2_ftps_down_1a.csv};
            \addlegendentry{\cite{9320384}}

            others: black with different line styles
            \addplot[kit-orange100, loosely dashed, thick]
                table [x index=0, y index=2, col sep=comma]
                {data/payload_cdf_log_necstgen_v3_ftps_down_1a.csv};
            \addlegendentry{\cite{10000731}}

            \addplot[black, dash pattern=on 8pt off 3pt, thick]
                table [x index=0, y index=2, col sep=comma]
                {data/payload_cdf_log_trafficgpt10_ftps_down_1a.csv};
            \addlegendentry{\cite{qu2024trafficgpt} (Adapted)}

            \addplot[black, densely dotted, thick]
                table [x index=0, y index=2, col sep=comma]
                {data/payload_cdf_log_icetran_v2_ftps_down_1a.csv};
            \addlegendentry{\cite{9145646}}

            \addplot[kit-green100, dashdotdotted, thick]
                table [x index=0, y index=2, col sep=comma]
                {data/payload_cdf_log_timegan_v2_ftps_down_1a.csv};
            \addlegendentry{\cite{NEURIPS2019_c9efe5f2}}

            \addplot[kit-cyan100, densely dashed, very thick]
                table [x index=0, y index=2, col sep=comma]
                {data/payload_cdf_log_algo_v3_3_ftps_down_1a.csv};
            \addlegendentry{Proposed}
            \end{axis}
        \end{tikzpicture}
        \caption{Payload length (bytes, log scaled)}
        \label{fig:cdf_payload_comp_ftps}
    \end{subfigure}
    \hfill
    \begin{subfigure}[t]{\columnwidth}
        \centering
        \begin{tikzpicture}
            \begin{axis}[xmode=log, filter discard warning=false, unbounded coords=discard,
                width=\cdfaxiswidth,
                height=0.4\linewidth,
                scale only axis,
                grid style={line width=.2pt, gray!50},
                major grid style={line width=.3pt, gray!60},
                minor tick num=1,
                scaled x ticks=false,
                xticklabel style={
                    /pgf/number format/fixed,
                    /pgf/number format/precision=3
                },
                tick label style={font=\small},
                ylabel style={font=\small},
                xlabel style={font=\small},
                xlabel={},
                ylabel={CDF},
                xmin=0.006,
                ymin=0.0, ymax=1.01,
                grid=both,
                legend cell align={left},
                legend pos=south east,
                legend style={font=\footnotesize, draw=black, fill=white, fill opacity=.8},
            ]
            \addplot[kit-purple100, solid, thick]
                table [x index=0, y index=1, col sep=comma]
                {data/time_cdf_log_algo_v3_3_ftps_down_1a.csv};
            \addlegendentry{Real}

            \addplot[black, dashdotted, thick]
                table [x index=0, y index=2, col sep=comma]
                {data/time_cdf_log_gen_iot_v2_ftps_down_1a.csv};
            \addlegendentry{\cite{9320384}}

            \addplot[kit-orange100, loosely dashed, thick]
                table [x index=0, y index=2, col sep=comma]
                {data/time_cdf_log_necstgen_v3_ftps_down_1a.csv};
            \addlegendentry{\cite{10000731}}

            \addplot[black, dash pattern=on 8pt off 3pt, thick]
                table [x index=0, y index=2, col sep=comma]
                {data/time_cdf_log_trafficgpt10_ftps_down_1a.csv};
            \addlegendentry{\cite{qu2024trafficgpt} (Adapted)}

            \addplot[black, densely dotted, thick]
                table [x index=0, y index=2, col sep=comma]
                {data/time_cdf_log_icetran_v2_ftps_down_1a.csv};
            \addlegendentry{\cite{9145646}}

            \addplot[kit-green100, dashdotdotted, thick]
                table [x index=0, y index=2, col sep=comma]
                {data/time_cdf_log_timegan_v2_ftps_down_1a.csv};
            \addlegendentry{\cite{NEURIPS2019_c9efe5f2}}

            \addplot[kit-cyan100, densely dashed, very thick]
                table [x index=0, y index=2, col sep=comma]
                {data/time_cdf_log_algo_v3_3_ftps_down_1a.csv};
            \addlegendentry{Proposed}
            \end{axis}
        \end{tikzpicture}
        \caption{\ac{IAT} (seconds, log scaled)}
        \label{fig:cdf_iat_comp_ftps}
    \end{subfigure}

    \caption{Average per-flow \ac{CDF} comparison of payload length and \ac{IAT} for \ac{FTPS} traffic.}
    \label{fig:cdf_payload_iat_ftps}
\end{figure}

\textcolor{red}{The \ac{FTPS} in Fig.~\ref{fig:cdf_payload_iat_ftps} and Facebook video in Fig.~\ref{fig:cdf_payload_iat_facebook_video} traces present the most severe heavy-tailed challenges, combining sub-millisecond hardware bursts with multi-second application idle times. On the Facebook Video trace, the proposed generator overcomes the mode collapse problem of recurrent and continuous baselines (e.g., \cite{9320384} and \cite{NEURIPS2019_c9efe5f2}), successfully tracking the true \ac{IAT} \ac{CDF} across multiple orders of magnitude without truncating the tail. However, the \ac{FTPS} trace reveals a structural limitation. While the proposed model achieves excellent $\bm{\mathcal{D}}_{\mathrm{div}}$ scores on \ac{FTPS}, its average per-flow \ac{IAT} \ac{CDF} exhibits mode collapse. This discrepancy between global and per-flow metrics indicates that while the \ac{MDN} successfully learned the overall geometric volume of the heavy tail, it struggled to distribute these extreme \acp{IAT} correctly across flows of varying lengths. An analysis of gaps exceeding \SI{1.0}{\sec} reveals that in the real \ac{FTPS} trace, $74.3\%$ of extreme \acp{IAT} occur within short to medium flows ($\le 100$ packets). Conversely, the synthetic model allocated $97.9\%$ of its extreme gaps to long bulk flows ($> 100$ packets). This inversion suggests that the flow-length conditioning mechanism over-associated the idle state with bulk data transfers, likely because long flows dominate the packet-level training batches. This provides a future direction to further improve the model. }

\begin{table*}[t]
\centering
\caption{Ablation study evaluating the impact of architectural components on Facebook video traffic.}
\label{tab:ablation_results}
\resizebox{\textwidth}{!}{%
\begin{tabular}{lcccccccccc}
\hline
\multirow{2}{*}{Model Variant} & \multicolumn{2}{c}{$\mathcal{D}_{\mathrm{temp}}$ (\ac{AC})} & \multicolumn{2}{c}{\ac{KS} Statistic} & \multicolumn{2}{c}{Log-\ac{WD}}   & \multicolumn{2}{c}{Linear-\ac{WD}}  & \multicolumn{2}{c}{Tail Extremes}    \\ \cline{2-11} 
                               & Payload                      & IAT                          & Payload           & IAT               & Payload         & IAT             & Payload           & IAT             & Max IAT (s)       & 99.9\% IAT (s)   \\ \hline
Full Model                     & $0.12$                       & $0.12$                       & $\mathbf{0.248}$  & $0.353$           & $\mathbf{0.69}$ & $0.43$          & $\mathbf{215.11}$ & $8.92$          & $\mathbf{106.68}$ & $\mathbf{82.55}$ \\
Variant A (No Idle)            & $0.13$                       & $0.12$                       & $0.423$           & $0.402$           & $1.18$          & $0.49$          & $521.25$          & $9.42$          & $25.20$           & $17.11$          \\
Variant B (No Flow Len)        & $\mathbf{0.10}$              & $\mathbf{0.09}$              & $0.613$           & $0.426$           & $2.05$          & $0.50$          & $511.37$          & $9.50$          & $0.04$            & $0.04$           \\
Variant C (Gaussian)           & $0.12$                       & $0.12$                       & $0.285$           & $\mathbf{0.316}$  & $0.97$          & $\mathbf{0.23}$ & $547.82$          & $\mathbf{8.16}$ & $64.99$           & $59.43$          \\
Variant D (Baseline)           & $\mathbf{0.10}$              & $0.12$                       & $0.601$           & $0.405$           & $2.35$          & $0.50$          & $618.02$          & $9.50$          & $0.12$            & $0.11$           \\ \hline
\textit{Real Data (Targets)}   & \textit{-}                   & \textit{-}                   & \textit{-}        & \textit{-}        & \textit{-}      & \textit{-}      & \textit{-}        & \textit{-}      & \textit{278.81}   & \textit{278.80}  \\ \hline
\end{tabular}%
}
\end{table*}

\textcolor{red}{Taking all datasets into account, the proposed hybrid architecture proves to be highly effective. Despite the localized distribution error in \ac{FTPS}, and while it does not strictly outperform the best baseline in every single metric, the visual and quantitative results confirm its overall success. The model consistently provides a robust, high-fidelity approximation of complex traffic structures across diverse applications, while retaining unique advantages in interpretability and maintaining a low memory footprint.}

\textcolor{red}{
\subsection{Ablation study}
To isolate the contribution of the main design choices, we evaluate the proposed generator on the Facebook video trace with four ablated variants. The full model combines the idle state, flow-length conditioning, and Student-$t$ emissions. Variant A removes the idle state. Variant B removes flow-length conditioning. Variant C keeps the same architecture but replaces Student-$t$ emissions with Gaussian emissions. Variant D is the baseline that removes both the idle state and flow-length conditioning and also uses Gaussian emissions. We focus on Facebook video because it is one of the more challenging datasets in our study. It exhibits strong heavy-tail behavior in both payload length and \ac{IAT}, yet introduces a manageable computational overhead. Table~\ref{tab:ablation_results} reports the quantitative comparison.}

\textcolor{red}{The results show that the full model provides the best overall balance across payload fidelity, temporal behavior, and tail coverage. In particular, it yields the lowest payload discrepancy among all variants in terms of \ac{KS} and \ac{WD} metrics, with $\mathrm{KS}_{\mathrm{PL}}=0.248$, $\mathrm{WD}_{\mathrm{PL}}^{\log}=0.69$, and $\mathrm{WD}_{\mathrm{PL}}^{\mathrm{lin}}=215.11$. 
}

\textcolor{red}{Among the ablations, removing the flow-length feature causes the clearest degradation. Variant B achieves slightly smaller \ac{AC} errors, but its distributional fidelity deteriorates strongly, with $\mathrm{KS}_{\mathrm{PL}}=0.613$ and $\mathrm{KS}_{\mathrm{IAT}}=0.426$. More importantly, its generated tail collapses almost completely, with a maximum \ac{IAT} of only \SI{0.04}{\second}. 
This shows that flow length is a necessary conditioning signal that helps the generator distinguish short flows from long ones.}

\textcolor{red}{Removing the idle state also degrades the performance. Variant A increases the payload mismatch to $\mathrm{KS}_{\mathrm{PL}}=0.423$ and $\mathrm{WD}_{\mathrm{PL}}^{\mathrm{lin}}=521.25$, while reducing the maximum generated \ac{IAT} from $106.68$ s to $25.20$ s. 
The result supports the role of the idle state in representing rare but operationally important waiting periods.}

\textcolor{red}{The comparison between the full model and Variant C clarifies the effect of the Student-$t$ emission family. Variant C attains slightly better bulk-oriented \ac{IAT} scores, with $\mathrm{KS}_{\mathrm{IAT}}=0.316$, $\mathrm{WD}_{\mathrm{IAT}}^{\log}=0.23$, and $\mathrm{WD}_{\mathrm{IAT}}^{\mathrm{lin}}=8.16$, whereas the full model gives $0.353$, $0.43$, and $8.92$, respectively. However, Variant C underestimates the far tail more strongly, causing a maximum \ac{IAT} of \SI{64.99}{\second} and a $99.9$th percentile \ac{IAT} of \SI{59.43}{\second}, compared with \SI{106.68}{\second} and \SI{82.55}{\second} for the full model. Therefore, the Gaussian emission is slightly better at matching the dense central part of the \ac{IAT} distribution, but the Student-$t$ emission preserves rare long gaps more faithfully. Since these long gaps correspond to idle periods that affect timeout, buffer draining, and state dwell behavior in a \ac{DT}, we prefer the Student-$t$ model despite the small advantage of Variant C on some average \ac{IAT} metrics.}

\textcolor{red}{Finally, Variant D performs worst overall. Its payload mismatch remains high, with $\mathrm{KS}_{\mathrm{PL}}=0.601$ and $\mathrm{WD}_{\mathrm{PL}}^{\log}=2.35$, and its \ac{IAT} tail almost disappears, with a maximum generated \ac{IAT} of only \SI{0.12}{\second}. This confirms that the gains do not come from a single modification alone. Instead, the performance of the proposed generator is obtained by the interaction of three complementary design choices. The fact that each ablation fails in a different and interpretable way also supports the modular structure of the generator, where the \ac{HMM} and the \ac{MDN} contribute distinct functions to the generated traffic.}


\begin{figure}[t]
    \centering
    \includegraphics[width=\linewidth]{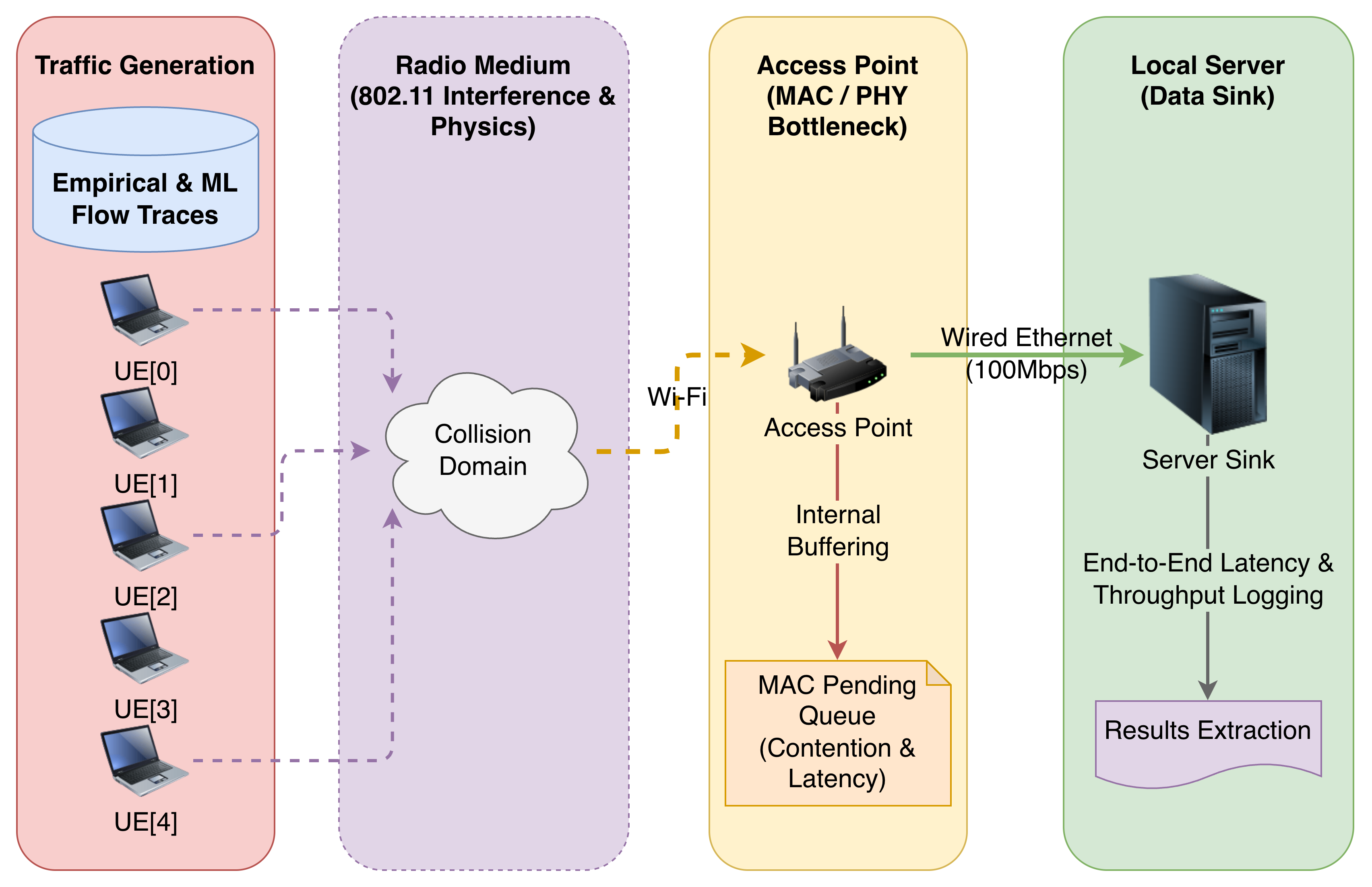}
    \caption{Omnet++ simulation setup.}
    \label{fig:omnet}
\end{figure}

\textcolor{red}{
\subsection{System-level RAN validation}
The previous results evaluate the generated traces with offline statistical metrics. To verify that these differences also matter at the system level, we inject the real traces, the proposed synthetic traces, and a Poisson baseline into an OMNeT++/INET wireless simulator and measure the resulting throughput, queue occupancy, and end-to-end latency \cite{varga2008omnetpp}. The topology is shown in Fig.~\ref{fig:omnet}. It contains five \ac{UE} nodes that contend for the same IEEE 802.11ac channel, a single access point, and a server. The real and proposed traffic are replayed from timestamped packet traces using a custom script, while the Poisson baseline is produced by \texttt{UdpBasicApp} \cite{varga2008omnetpp}. To isolate the effect of burst structure from the mean offered load, all three scenarios offer an identical aggregate application-layer volume of approximately \SI{950}{\mega\byte}. For the Poisson baseline, we achieve this by fitting the mean packet size and exponential \acp{IAT} to the specific data rate of each \ac{UE}.
}

\textcolor{red}{
For the real trace, the dataset is partitioned so each of the five \acp{UE} receives same number of distinct subset of flows. To evaluate network contention across $30$ simulation runs, the flow order within each \ac{UE} is randomly shuffled per run. This preserves the packet-level burst structure inside each flow while randomizing collisions between users. The proposed model generates synthetic flows independently for each \ac{UE} to match the exact byte budget.
}

\textcolor{red}{
Fig.~\ref{fig:throughput_combined} compares the received throughput time series for a particular run. Even though all three scenarios offer the same aggregate data volume, their short-term behavior and delivered throughput vary significantly. The Poisson baseline distributes traffic uniformly. This allows the \ac{MAC} layer to process packets smoothly with almost no drops, resulting in \SI{946}{\mega\byte} of successfully delivered data. The real trace, in contrast, contains long quiet periods followed by strong macro-bursts. These heavy-tailed bursts overwhelm the finite \ac{MAC} layer buffers and cause severe packet drops, reducing the total delivered volume to roughly \SI{310}{\mega\byte}. The proposed generator reproduces this burst-dominated pattern and the resulting network congestion much more closely, yielding an aggregate delivered volume of approximately \SI{500}{\mega\byte}. This shows that scheduling and buffer dynamics are driven by bursty periods rather than long-term averages.
}

\begin{figure}[t]
    \centering
    \begin{tikzpicture}
        \begin{axis}[
            width=\cdfaxiswidth,
            height=0.5\linewidth,
            scale only axis,
            grid style={line width=.2pt, gray!50},
            major grid style={line width=.3pt, gray!60},
            minor tick num=1,
            tick label style={font=\small},
            ylabel style={font=\small},
            xlabel style={font=\small},
            xlabel={Simulation Time (minutes)},
            ylabel={Throughput (Kbps, log scale)},
            xmin=0, xmax=185,
            ymin=1,
            ymode=log,
            grid=both,
            legend cell align={left},
            legend pos=north east,
            legend style={font=\scriptsize, draw=black, fill=white, fill opacity=.9},
            every axis plot/.append style={very thin}
        ]
        
        \addplot[black, solid, thick] 
            table [x expr=\thisrow{time}/60, y expr={max(1, \thisrow{kbps})}] {data/throughput_basic.dat};
        \addlegendentry{Poisson}
        
        \addplot[kit-purple100, ycomb, thick, opacity=0.8, forget plot] 
            table [x expr=\thisrow{time}/60, y expr={max(1, \thisrow{kbps})}] {data/throughput_real.dat};
        \addlegendimage{kit-purple100, solid, thick, opacity=0.8}
        \addlegendentry{Real Trace}
        
        \addplot[kit-cyan100, ycomb, thick, opacity=0.6, forget plot] 
            table [x expr=\thisrow{time}/60, y expr={max(1, \thisrow{kbps})}] {data/throughput_ml.dat};
        \addlegendimage{kit-cyan100, solid, thick, opacity=0.6}
        \addlegendentry{Proposed}
        
        \end{axis}
    \end{tikzpicture}
    \caption{\textcolor{red}{Combined throughput time series under equal average offered load.}}
    \label{fig:throughput_combined}
\end{figure}

\textcolor{red}{
The system-level consequence of this contention is visible in Fig.~\ref{fig:system_level_validation}, which reports the empirical \acp{CDF} of \ac{MAC} queue size and end-to-end latency averaged over the simulation runs. The Poisson baseline keeps the \ac{MAC} queue small for most packets and underestimates the latency penalty. Under the real trace, the queue size builds up and frequently hits the maximum \ac{MAC} buffer capacity of approximately \SI{1}{\mega\byte}, which aligns with the high packet drop rate observed in the throughput analysis. The latency CDF shifts to the right due to these saturated buffers. The proposed generator closely tracks the real trace in both metrics, capturing the queue saturation limit and the long tail of the latency distribution. Hence, the gains observed in the offline packet-level evaluation translate into realistic system-level stress once the traffic interacts with a shared wireless channel and a full protocol stack. This addresses the digital-twin requirement that synthetic traffic must induce realistic queueing and latency behavior in network simulations.
}

\begin{figure}[t]
    \centering
    \begin{tikzpicture}
        
        \begin{axis}[
            width=\cdfaxiswidth,
            height=0.5\linewidth,
            scale only axis,
            axis x line*=bottom, 
            axis y line*=left,   
            ylabel shift={-2pt}, 
            xmin=-0.001, xmax=0.9,
            ymin=0.0, ymax=1.04,
            xlabel={MAC Queue Size (\SI{}{\mega\byte})},
            ylabel={CDF},
            tick label style={font=\small},
            ylabel style={font=\small},
            xlabel style={font=\small},
            ymajorgrids=true, 
            grid style={line width=.1pt, gray!50},
            ytick distance=0.1, 
        ]
        
        \addplot[kit-green, solid, very thick]
            table [x=queue_mb, y=cdf] {data/queue_mb_cdf_basic.dat};

        \addplot[kit-purple100, solid, very thick, opacity=0.9]
            table [x=queue_mb, y=cdf] {data/queue_mb_cdf_real.dat};

        \addplot[kit-cyan100, solid, very thick, opacity=0.9]
            table [x=queue_mb, y=cdf] {data/queue_mb_cdf_ml.dat};

        \end{axis}
        
        \begin{axis}[
            width=\cdfaxiswidth,
            height=0.5\linewidth,
            scale only axis,
            axis x line*=top,
            axis y line*=right,  
            ytick=\empty,        
            xmode=log,
            xmin=1e-1, xmax=4e2, 
            ymin=0.0, ymax=1.04,
            xlabel={End-to-End Latency (ms, log scale)},
            tick label style={font=\small},
            xlabel style={font=\small},
            legend columns=1,
            legend cell align={left},
            legend style={
                at={(0.35,0.5)},anchor=center,
                font=\scriptsize, 
                draw=black, 
                fill=white, 
                fill opacity=.80,
                text opacity=1 
            },
        ]

        \addlegendimage{kit-green, solid, very thick}
        \addlegendentry{Queue: Poisson}
        \addlegendimage{kit-purple100, solid, very thick}
        \addlegendentry{Queue: Real}
        \addlegendimage{kit-cyan100, solid, very thick}
        \addlegendentry{Queue: Prop.}

        \addlegendimage{kit-green, dashed, very thick}
        \addlegendentry{Lat: Poisson}
        \addlegendimage{kit-purple100, dashed, very thick}
        \addlegendentry{Lat: Real}
        \addlegendimage{kit-cyan100, dashed, very thick}
        \addlegendentry{Lat: Prop.}

        \addplot[forget plot, kit-green, dashed, very thick]
            table [x=latency, y=cdf] {data/latency_cdf_basic.dat};

        \addplot[forget plot, kit-purple100, dashed, very thick, opacity=0.9]
            table [x=latency, y=cdf] {data/latency_cdf_real.dat};

        \addplot[forget plot, kit-cyan100, dashed, very thick, opacity=0.9]
            table [x=latency, y=cdf] {data/latency_cdf_ml.dat};

        \end{axis}
    \end{tikzpicture}
    
    \caption{\textcolor{red}{System-level RAN stress comparison. Solid lines show the \ac{CDF} of MAC queue sizes, while the dashed curves show the end-to-end latency \ac{CDF}.}}
    \label{fig:system_level_validation}
\end{figure}

\section{Conclusion}
In this article we presented a compact packet level traffic generator for network \acp{DT} that couples a small \ac{HMM} with a Student t mixture density emission model to handle heavy tailed payloads and \acp{IAT}. \textcolor{red}{The tail anchored idle state and flow aware conditioning allow the model to reproduce buffering, streaming, and idle phases together with realistic joint distributions of payload sizes and \acp{IAT} spanning nearly two orders of magnitude for payloads (e.g., $40$ to \SI{1550}{\byte}) and over six orders of magnitude for inter-arrival times (e.g., sub-milliseconds to tens of seconds).} \textcolor{red}{Across the four datasets used in this investigation,} it matches marginal distributions, temporal autocorrelation, and flow diversity more closely than recent neural, transformer, and \ac{HMM} baselines in most cases, while keeping a memory footprint in the order of a few tenths of a MB. \textcolor{red}{The separation between \ac{HMM} and \ac{MDN} makes the generator interpretable, which, combined with its compact size, supports practical deployment inside resource constrained network \acp{DT}.}


\bibliographystyle{IEEEtran}
\bibliography{ref}

@ARTICLE{18626,
  author={Rabiner, L.R.},
  journal={Proc. IEEE}, 
  title={A tutorial on hidden Markov models and selected applications in speech recognition}, 
  year={1989},
  volume={77},
  number={2},
  pages={257-286},
  doi={10.1109/5.18626}}

@misc{hmmlearn,
  title = {hmmlearn: Hidden Markov Models in Python},
  author = {Cournapeau, David and Pedregosa, Fabian and Varoquaux, Gael and Lebedev, Sergei and Lee, Antony and Danielson, Matthew},
  year = {2015},
  publisher = {GitHub},
  journal = {GitHub repository},
  url = {https://github.com/hmmlearn/hmmlearn},
  note = {{accessed Nov. 12, 2025}}
}

@ARTICLE{282603,
  author={Leland, W.E. and Taqqu, M.S. and Willinger, W. and Wilson, D.V.},
  journal={IEEE/ACM Trans. Netw.}, 
  title={On the self-similar nature of Ethernet traffic (extended version)}, 
  year={1994},
  volume={2},
  number={1},
  pages={1-15},
  doi={10.1109/90.282603}}

@ARTICLE{650143,
  author={Crovella, M.E. and Bestavros, A.},
  journal={IEEE/ACM Trans. Netw.}, 
  title={Self-similarity in World Wide Web traffic: evidence and possible causes}, 
  year={1997},
  volume={5},
  number={6},
  pages={835-846},
  doi={10.1109/90.650143}}

@INPROCEEDINGS{koktas2025state,
  author={Koktas, Enes and Rost, Peter},
  booktitle={2025 IEEE 36th IEEE Int. Symp. Pers. Indoor Mob. Radio Commun. (PIMRC)}, 
  title={State Aware Traffic Generation for Real-Time Network Digital Twins}, 
  year={2025},
  pages={1-6},
  doi={10.1109/PIMRC62392.2025.11274598}}

@article{student_t_ref,
 URL = {https://doi.org/10.1080/01621459.1989.10478852},
 author = {Kenneth L. Lange and Roderick J. A. Little and Jeremy M. G. Taylor},
 journal = {J. Amer. Statist. Assoc.},
 number = {408},
 pages = {881--896},
 publisher = {[American Statistical Association, Taylor & Francis, Ltd.]},
 title = {Robust Statistical Modeling Using the t Distribution},
 urldate = {2025-11-18},
 volume = {84},
 year = {1989}
}

@ARTICLE{9899718,
  author={Mihai, Stefan and Yaqoob, Mahnoor and Hung, Dang V. and Davis, William and Towakel, Praveer and Raza, Mohsin and Karamanoglu, Mehmet and Barn, Balbir and Shetve, Dattaprasad and Prasad, Raja V. and Venkataraman, Hrishikesh and Trestian, Ramona and Nguyen, Huan X.},
  journal={IEEE Commun. Surveys Tuts.}, 
  title={Digital Twins: A Survey on Enabling Technologies, Challenges, Trends and Future Prospects}, 
  year={2022},
  volume={24},
  number={4},
  pages={2255-2291},
  doi={10.1109/COMST.2022.3208773}}

@techreport{irtf-nmrg,
    institution =   {Internet Engineering Task Force},
    publisher = {Internet Engineering Task Force},
    note =      {Work in Progress},
    url =       {https://datatracker.ietf.org/doc/draft-irtf-nmrg-network-digital-twin-arch/05/},
    author =    {Cheng Zhou and Hongwei Yang and Xiaodong Duan and Diego Lopez and Antonio Pastor and Qin Wu and Mohamed Boucadair and Christian Jacquenet},
    title =     {Network Digital Twin: Concepts and Reference Architecture},
    pagetotal = 25,
    year =      2024,
    month =     mar
}

@INPROCEEDINGS{9320384,
  author={Shahid, Mustafizur R. and Blanc, Gregory and Jmila, Houda and Zhang, Zonghua and Debar, Hervé},
  booktitle={2020 IEEE 25th Pacific Rim Int. Symp. on Dependable Comput. (PRDC)}, 
  title={Generative Deep Learning for Internet of Things Network Traffic Generation}, 
  year={2020},
  volume={},
  number={},
  pages={70-79},
  doi={10.1109/PRDC50213.2020.00018}
}

@INPROCEEDINGS{9145646,
  author={Red{\v{z}}ovi{\'c}, Hasan and Smiljani{\'c}, Aleksandra and Bjelica, Milan},
  booktitle={Proc. of 4th Int. Conf. on Electr., Electron. and Comput. Eng. (IcETRAN)}, 
  title={{IP} Traffic Generator Based on Hidden {M}arkov Models}, 
  year={2017},
  month={Jun},
  pages={TEI2.3.1-6},
  isbn={978-86-7466-692-0}
}

@article{3664655,
author = {Li, Tong and Hui, Shuodi and Zhang, Shiyuan and Wang, Huandong and Zhang, Yuheng and Hui, Pan and Jin, Depeng and Li, Yong},
title = {Mobile User Traffic Generation Via Multi-Scale Hierarchical {GAN}},
year = {2024},
issue_date = {September 2024},
publisher = {Association for Computing Machinery},
address = {New York, NY, USA},
volume = {18},
number = {8},
issn = {1556-4681},
url = {https://doi.org/10.1145/3664655},
doi = {10.1145/3664655},
journal = {ACM Trans. Knowl. Discov. Data},
month = jul,
articleno = {189},
numpages = {19}
}

@inproceedings{NEURIPS2019_c9efe5f2,
 author = {Yoon, Jinsung and Jarrett, Daniel and van der Schaar, Mihaela},
 booktitle = {Adv. in Neural Inf. Process. Syst.},
 editor = {H. Wallach and H. Larochelle and A. Beygelzimer and F. d\textquotesingle Alch\'{e}-Buc and E. Fox and R. Garnett},
 pages = {},
 publisher = {Curran Associates, Inc.},
 title = {Time-series Generative Adversarial Networks},
 url = {https://proceedings.neurips.cc/paper_files/paper/2019/file/c9efe5f26cd17ba6216bbe2a7d26d490-Paper.pdf},
 volume = {32},
 year = {2019}
}

@INPROCEEDINGS{10000731,
  author={Meslet-Millet, Fabien and Mouysset, Sandrine and Chaput, Emmanuel},
  booktitle={IEEE Global Commun. Conf. (GLOBECOM)}, 
  title={{NeCSTGen}: An approach for realistic network traffic generation using Deep Learning}, 
  year={2022},
  volume={},
  number={},
  pages={3108-3113},
  doi={10.1109/GLOBECOM48099.2022.10000731}
}

@misc{qu2024trafficgpt,
      title={{TrafficGPT}: Breaking the Token Barrier for Efficient Long Traffic Analysis and Generation}, 
      author={Jian Qu and Xiaobo Ma and Jianfeng Li},
      year={2024},
      eprint={2403.05822},
      archivePrefix={arXiv},
      primaryClass={cs.LG},
      url={https://arxiv.org/abs/2403.05822}, 
}

@inproceedings{3673792,
author = {Chu, Andrew and Jiang, Xi and Liu, Shinan and Bhagoji, Arjun and Bronzino, Francesco and Schmitt, Paul and Feamster, Nick},
title = {Feasibility of State Space Models for Network Traffic Generation},
year = {2024},
isbn = {9798400707131},
publisher = {Association for Computing Machinery},
address = {New York, NY, USA},
url = {https://doi.org/10.1145/3672198.3673792},
doi = {10.1145/3672198.3673792},
booktitle = {Proc. of the 2024 SIGCOMM Workshop on Networks for AI Comput.},
pages = {9–17},
numpages = {9},
location = {Sydney, NSW, Australia},
series = {NAIC '24}
}

@inproceedings{3423643,
author = {Lin, Zinan and Jain, Alankar and Wang, Chen and Fanti, Giulia and Sekar, Vyas},
title = {Using {GANs} for Sharing Networked Time Series Data: Challenges, Initial Promise, and Open Questions},
year = {2020},
publisher = {Association for Computing Machinery},
address = {New York, NY, USA},
url = {https://doi.org/10.1145/3419394.3423643},
doi = {10.1145/3419394.3423643},
booktitle = {Proc. of the ACM Internet Meas. Conf.},
pages = {464–483},
numpages = {20},
series = {IMC '20}
}

@inproceedings{draper2016,
  title={Characterization of encrypted and {VPN} traffic using time-related},
  author={Draper-Gil, Gerard and Lashkari, Arash Habibi and Mamun, Mohammad Saiful Islam and Ghorbani, Ali A},
  booktitle={Proc. of the 2nd Int. Conf. on Inf. Syst. Secur. and Privacy (ICISSP)},
  pages={407--414},
  year={2016}
}

@ARTICLE{paxson1995poisson,
  author={Paxson, V. and Floyd, S.},
  journal={IEEE/ACM Trans. on Netw.}, 
  title={Wide area traffic: the failure of Poisson modeling}, 
  year={1995},
  volume={3},
  number={3},
  pages={226-244},
  doi={10.1109/90.392383}}

@article{netdiffusion,
  title={Netdiffusion: Network data augmentation through protocol-constrained traffic generation},
  author={Jiang, Xi and Liu, Shinan and Gember-Jacobson, Aaron and Bhagoji, Arjun Nitin and Schmitt, Paul and Bronzino, Francesco and Feamster, Nick},
  journal={Proc. ACM Meas. Anal. Comput. Syst.},
  volume={8},
  number={1},
  pages={1--32},
  year={2024},
  publisher={ACM New York, NY, USA}
}

@ARTICLE{stouter,
  author={Liu, Xiaosi and Xu, Xiaowen and Liu, Zhidan and Li, Zhenjiang and Wu, Kaishun},
  journal={IEEE Trans. Mob. Comput.}, 
  title={Spatio-Temporal Diffusion Model for Cellular Traffic Generation}, 
  year={2026},
  volume={25},
  number={1},
  pages={257-271}
  }

@misc{arjovsky2017wassersteingan,
      title={Wasserstein GAN}, 
      author={Martin Arjovsky and Soumith Chintala and Léon Bottou},
      year={2017},
      eprint={1701.07875},
      archivePrefix={arXiv},
      primaryClass={stat.ML},
      url={https://arxiv.org/abs/1701.07875}, 
}

@inproceedings{varga2008omnetpp,
  author    = {Varga, Andr\'{a}s and Hornig, Rudolf},
  title     = {An Overview of the OMNeT++ Simulation Environment},
  booktitle = {Proc. 1st Int. Conf. on Sim. Tools and Tech. for Commun., Netw. and Syst.},
  series    = {SimuTools '08},
  year      = {2008},
  isbn      = {978-963-9799-20-2},
  location  = {Marseille, France},
  pages     = {1--10},
  doi       = {10.1145/1416222.1416290},
  address   = {ICST, Brussels, Belgium}
}

\end{document}